\documentclass[preprint2]{aastex}

\usepackage{psfig}

\newcommand{\msun}{\,\hbox{$M_{\odot}$}}
\newcommand{\lsun}{\,\hbox{$L_{\odot}$}}
\newcommand{\kms}{\,\hbox{\hbox{km}\,\hbox{s}$^{-1}$}}
\newcommand{\vrot}{\,\hbox{$V_{\rm rot}$}}
\newcommand{\mbh}{\,\hbox{$M_{\rm BH}$}}
\newcommand{\msigma}{\,\hbox{$M_{\rm BH}-\sigma$}}
\newcommand{\pone}{Paper I}
\newcommand{\ptwo}{Paper II}
\newcommand{\degree}{\ensuremath{^\circ}}

\newcommand{\reff}{\mbox{$R_{\rm eff}$}}

\newcommand{\lir}{\,\hbox{$L_{\rm IR}$}}

\shorttitle{Host-galaxy dynamics and origin of PG QSOs}
\shortauthors{Dasyra et al.}

\begin{document}

\title{Host dynamics and origin of Palomar-Green QSOs \footnote{Based
on observations at the European Southern Observatory, Chile (171.B-0442)}}

\author{K. M. Dasyra\altaffilmark{2,3}, L. J. Tacconi\altaffilmark{2}, 
R. I. Davies\altaffilmark{2}, R. Genzel\altaffilmark{2}, 
D. Lutz\altaffilmark{2}, B. M. Peterson\altaffilmark{4},
S. Veilleux\altaffilmark{5}, A. J. Baker\altaffilmark{5,6,7}, 
M. Schweitzer\altaffilmark{2}, E. Sturm\altaffilmark{2}}

\altaffiltext{2}{Max-Planck-Institut f\"ur extraterrestrische Physik,
Postfach 1312, 85741, Garching, Germany}
\altaffiltext{3}{Spitzer Science Center, California Institute of Technology,
Mail Code 220-6, 1200 East California Blvd, Pasadena, CA 91125}
\altaffiltext{4}{Department of Astronomy, The Ohio State University, 140 West
18th Avenue, Columbus, OH 43210}
\altaffiltext{5}{Department of Astronomy, University of Maryland, College 
Park, MD 20742}
\altaffiltext{6}{Jansky Fellow, National Radio Astronomy Observatory}
\altaffiltext{7}{Department of Physics and Astronomy, Rutgers, the State 
University of New Jersey, 136 Frelinghuysen Road, Piscataway, NJ 08854}

\begin{abstract}
We present host-galaxy velocity dispersions of 12 local (mainly Palomar--Green)
quasi-stellar objects (QSOs) measured directly from the stellar CO absorption 
features in the $H$ band. The mean bulge velocity dispersion of the QSOs in our
sample is 186 \kms\ with a standard deviation of 24 \kms. The measurement of 
the stellar velocity dispersion in QSOs enables us to place them on 
observational 
diagrams such as the local black-hole mass to bulge-velocity-dispersion 
relation and the fundamental plane of early-type galaxies. Concerning the 
former relation, these QSOs have higher black hole masses than most Seyfert 
1 AGNs with similar velocity dispersions.
On the fundamental plane, PG QSOs are located between the regions occupied by  
moderate-mass and giant ellipticals. The QSO bulge and black hole 
masses, computed from the stellar velocity dispersions, are of order 
$10^{11}\msun$ and $10^8\msun$ respectively. The Eddington 
efficiency of their black holes is on average 0.25, assuming that all of 
the bolometric luminosity originates from the active nucleus. Our data
are consistent with other lines of evidence that Palomar--Green QSOs are 
related to galaxy mergers with gas-rich components and that they are 
formed in a manner similar to the most massive Ultraluminous Infrared 
Galaxies, regardless of their far-infrared emission. However, PG QSOs seem 
to have smaller host dispersions and different formation mechanisms than QSOs 
with supermassive black holes of $5 \times 10^8-10^9\msun$ that accrete at 
low rates and reside in massive spheroids.
\end{abstract}



\section{Introduction}
\label{sec:intro}

In the current picture of galaxy formation and evolution, starburst
and active-galactic-nucleus (AGN) activity are believed to be closely 
linked to each other and to merger events (e.g., \citealt{norman}; 
\citealt{haeh93}; \citealt{guinevere}; \citealt{dimatteo05}; 
\citealt{springel05}; \citealt{hopkins05}; \citealt{lipari06}). The common 
feature is the presence of gas, which is indispensible for initiating 
starbursts and feeding the AGN and which has been observed in local QSOs 
by \cite{sco03} and \cite{evans01}. 
However, the details of how mergers of gas-rich galaxies trigger either
type of activity are not well understood, with the uncertainties mainly 
originating from our insufficient knowledge and treatment of the interstellar 
medium physics. Of such a nature is the debated question (e.g., 
\citealt{joseph99}; \citealt{sanders99}) whether or not quasi-stellar-object 
(QSO) phases can be associated with Ultraluminous Infrared Galaxies 
(ULIRGs\footnote{Gas-rich mergers with infrared, 8-1000 \micron, luminosity 
greater than 10$^{12}$ \lsun.}; \citealt{sami96}). 

Various scenarios have been proposed that relate QSOs and ULIRGs. 
Sanders et al. (1998a,b) suggested that after the nuclei of two merging 
galaxies coalesce, the IR emission that arises from dust enshrouding
circumnuclear starbursts and AGN  is strong enough for the system to 
reach a QSO-like luminosity. Later in time, AGN winds and supernova 
feedback clear out the dust and gas from the nuclear region.
The system goes through an optically bright phase
before further accretion and star formation are finally terminated.
According to these authors, every ULIRG should eventually go through a QSO 
phase once its nucleus is revealed. It is, of course, possible that by the 
time the nuclear region is optically less obscured, the gas that remains
does not fuel the AGN at a rate sufficient to make the latter shine as bright
as a QSO. Therefore, a more plausible scenario is that some ULIRGs may evolve 
into  QSOs, depending on the amount of gas consumed during their early-merger 
stages. Recent models predict a short (up to ~10$^8$ yrs) QSO phase after 
the nuclear coalescence of gas-rich galaxies (\citealt{dimatteo05}; 
\citealt{springel05}; \citealt{hopkins05}; \citealt{cattaneo05}). The 
outcome of these simulations depends strongly on the treatment of the 
interstellar medium.

Another scenario is that QSOs and ULIRGs do not necessarily follow each other 
on some evolutionary sequence, but they are both triggered by similar 
conditions (e.g., \citealt{canalizo01}). This assumption could hold, for
example, if ULIRGs are triggered by major mergers and QSOs by minor mergers 
of gas-rich galaxies (e.g., \citealt{canalizo01}, \citealt{veilleux06}).

All these scenarios were initially motivated by the fact that ULIRGs have 
IR luminosities $>10^{12}\lsun$ and number densities comparable to those of 
local QSOs (\citealt{sanders88a}) in the Bright Quasar Survey (BQS; 
\citealt{pg}) within the Palomar-Green (PG) catalog. 
Furthermore, PG QSOs have on average ``warmer'' IR spectral energy 
distributions (SEDs) than ULIRGs, which agrees with potential evolutionary 
schemes (\citealt{sanders89}). Some of them have host galaxies with signs 
of recent interaction (e.g. \citealt{surace01}). 

\cite{canalizo01} have pointed out that the evolutionary link between QSOs and 
ULIRGs is best elucidated by detecting starbursts in QSO host galaxies with 
the aid of spectroscopy. In the optical regime, spectra of the QSO host galaxy 
(or of star-forming regions in it) have been successfully modelled by 
\cite{canalizo01} for the derivation of the stellar ages and populations.
Spectroscopy also provides crucial information about the stellar kinematics 
of the QSO host galaxy. However, this information is hard to extract from
optical spectra, where the dilution of the host light by
the AGN continuum is high. The best-suited wavelength regime for the 
extraction of the galaxy dynamics is the near-infrared (NIR) $H$ band. The 
ratio of the host galaxy to the QSO photon flux is at a maximum there, because 
the SED of many stellar populations has a maximum at $\sim$ 1.6 \micron\
and the AGN flux has a minimum at $\sim$ 1.2 \micron\ (\citealt{elvis94}).
The AGN flux
increases at shorter wavelengths due to accretion-disk power-law emission 
and at longer wavelengths due to thermal emission of hot dust. Host-galaxy 
dynamic studies were performed both in the $H$ and in the $K$ band by 
\cite{oliva95} and \cite{oliva99}. However, these studies mainly presented 
observations of Seyfert 1 AGNs with bolometric luminosities lower than those 
of QSOs. 

We aimed to investigate the possibility that QSOs and ULIRGs are related 
by comparing the dynamics of the stellar populations in local PG QSOs to 
those of ULIRGs. For this purpose, we have carried out a European 
Southern Observatory (ESO) large program\footnote{PI Tacconi} to acquire 
spectroscopic data for 54 ULIRGs (including sources presented by
\citealt{genzel01} and \citealt{tacconi02}) and 12 QSOs. We have presented
the ULIRG stellar kinematics in Dasyra et al. (2006a,b), hereafter Papers 
I and II. In this paper, we present the 
host dynamics of (mainly) PG QSOs. These Very Large Telescope (VLT)
spectroscopic data constitute part of a larger project that aims to 
determine whether local ULIRGs and QSOs can be related, through near- and 
mid- infrared spectroscopy and imaging of their hosts; the program is
called QUEST ({\it Quasar and ULIRG Evolutionary STudy}) and it is 
described in detail by \cite{veilleux06}.

This paper is arranged as follows.
We present our sample and briefly summarize the observations and data 
reduction techniques in \S~\ref{sec:obs}. The host and black hole (BH) 
properties of the QSOs in our sample, as inferred from the stellar 
dynamics, are given in \S~\ref{sec:results}; the dynamical results for
the PG QSOs are then compared to those of ULIRGs in \S~\ref{sec:selection}. 
We discuss the plausibility of a scenario that relates PG QSOs to gas-rich 
mergers and investigate the origin of local QSO populations in 
\S~\ref{sec:comp}. A summary is presented in \S~\ref{sec:conc}.


\section{Observations and data reduction}
\label{sec:obs}

\subsection{Sample selection}
\label{sec:sample}

To investigate the evolutionary scenarios associating ULIRGs with QSOs, we 
need to select amongst the local QSOs whose AGN properties most resemble 
those of ULIRGs. Since the AGNs in ULIRGs accrete at high rates (e.g., 
\ptwo), the local QSOs with the most active AGNs are ideal candidates.
The Bright QSO sample of the Palomar-Green catalog (\citealt{pg}) is  
well suited to this requirement; the point-like appearance, the U-B cutoff
selection, and, mainly, the $B$-band magnitude threshold ($\lesssim$16 mag) 
of these sources favors the selection of the most active local AGNs 
(\citealt{jester05}). 

The plethora of data available in the literature for PG QSOs allows for an 
optimal choice of a suitable sample. For example, measurements 
of the black-hole mass, \mbh, have been performed for 19 PG QSOs by 
\cite{kaspi00} and \cite{peterson04}. Approximately half of the sources 
in our sample were selected to have such a \mbh\ measurement. The remaining 
sources were chosen from the \cite{surace01} sample of so-called 
IR-excess QSOs. These are PG QSOs with strong mid- and far- infrared 
(MIR and FIR) emission compared to their optical luminosity; of the PG
sample, these are believed to most closely resemble ULIRGs. One more 
source, LBQS~0307-0101 (\citealt{chaffee}), was selected from the Large 
Bright QSO Survey, since it fits well within the observational constraints:
its $B$-band magnitude is similar to that of our entire sample. All the 
sources that we observed were at redshift $z \lesssim$ 0.1. This
upper redshift cutoff was selected to be smaller than that of the 
{\it Spitzer} subsample of QUEST, $\sim$0.3, to maximize the host signal 
and facilitate the extraction of the stellar kinematics. 
The identifiers, coordinatess, redshifts, and $B$-band magnitudes of the 12 
sources selected and observed are given in Table~\ref{tab:list}.

To investigate how well our PG subsample represents the entire BQS PG
population, we examined the optical and MIR/FIR luminosities of our sources 
as tabulated in Table~\ref{tab:lum}. The optical luminosities 
$\lambda L_{\lambda}(5100\dot{A})$ are from \cite{peterson04} and 
\cite{veste06}. We computed the infrared luminosities $L_{IR}$ from the 12, 
25, 60, and 100 \micron\ fluxes (\citealt{sanders89}; 
\citealt{IRASZ}; \citealt{haas03}) using the \cite{sami96} prescription.
The IR luminosity of all sources in our sample is greater than 
$10^{11}\lsun$. This luminosity threshold defines Luminous Infrared 
Galaxies (LIRGs). In Table~\ref{tab:lum}, we also present the ratio of 
the IR to optical/ultraviolet luminosity of our sources, as derived 
from \cite{surace01} and \cite{guyon06}. The latter is denoted as 
``big blue bump'' luminosity, $L_{\rm bbb}$, and it is computed by 
integrating the luminosity per unit wavelength from $\sim$0.01 to 1 \micron. 

\begin{figure*}
\centering
\includegraphics[width=8cm]{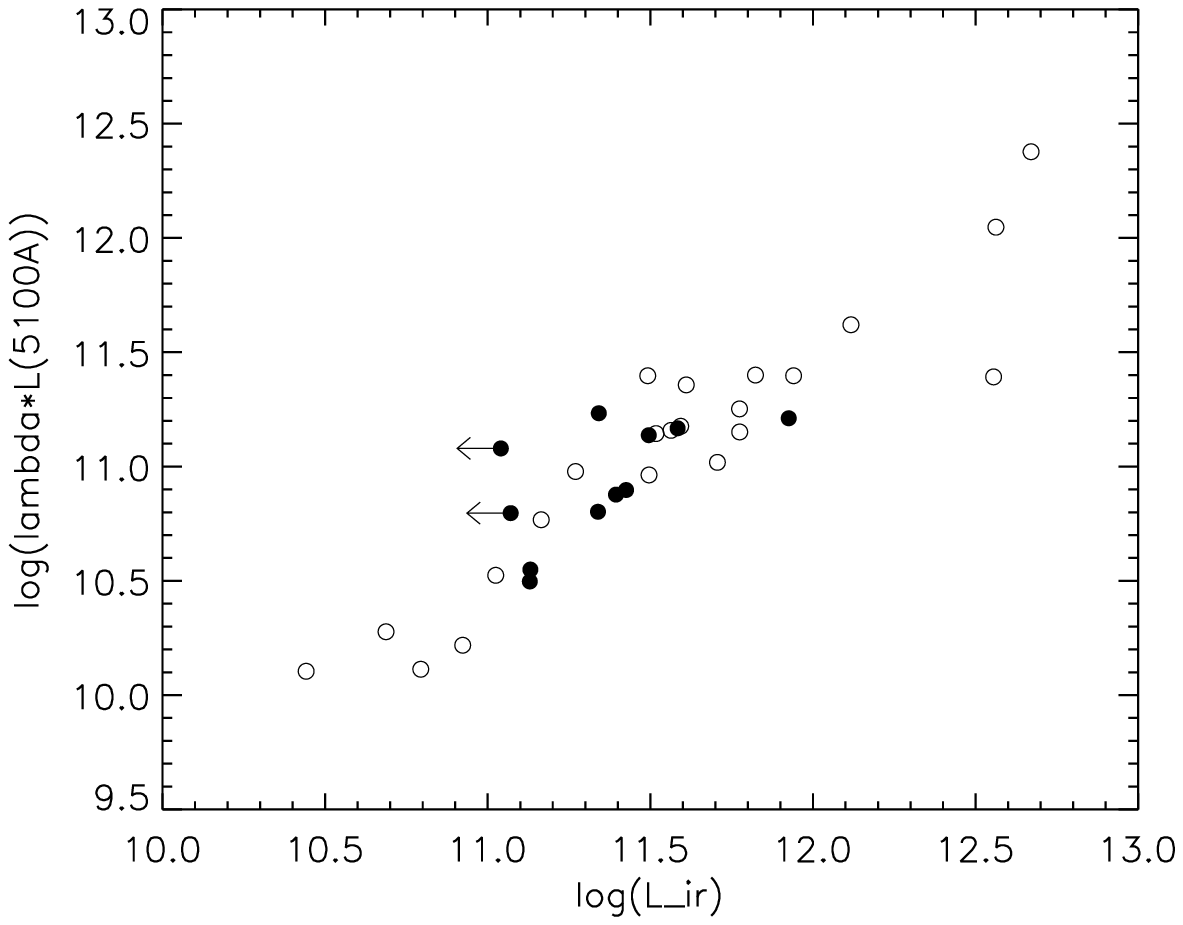}
\includegraphics[width=8cm]{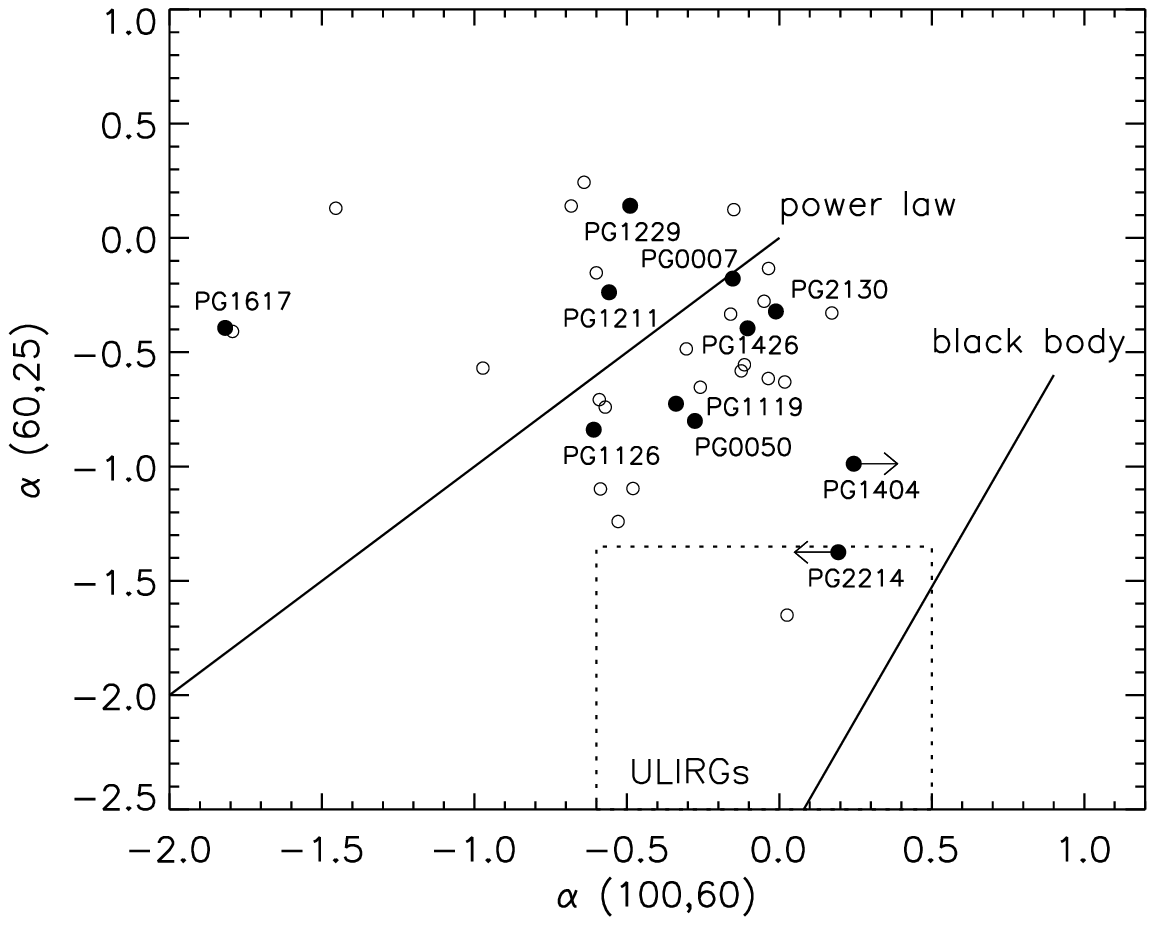}
\caption{{\it Left panel:} The optical vs. IR luminosity of PG QSOs. {\it
Right panel:} The MIR color-index diagram used to identify possible transition 
objects between ULIRGs and QSOs (see \citealt{lipari}; \citealt{canalizo01}). 
The empirically determined position of ULIRGs on this diagram is indicated by 
a dotted box (\citealt{canalizo01}). The power-law and black-body lines
roughly separate between AGN- and starburst- dominated sources. \newline
In both panels, the QSOs in our sample are plotted as filled circles. For
two QSOs in our sample with upper limits on their MIR fluxes, arrows indicate
how their position would change with a decrease of the respective flux. All 
other \cite{boroson} PG QSOs with exact MIR flux measurements are plotted as 
open circles. This diagram indicates that in terms of both optical and MIR/FIR 
properties, the QSOs in our sample are representative of the local PG 
population.
\label{fig:sample} }
\end{figure*}

In Fig.~\ref{fig:sample}, left panel, we plot the optical versus the
infrared luminosity of PG QSOs. To construct this diagram, we used 
PG QSOs in the entire QUEST redshift range. We also used those PG QSOs 
of the \cite{peterson04} and \cite{veste06} samples that have a flux 
measurement, instead of an upper limit, at 12, 25, 60, and 100 \micron. 
The sources with a mid- or far- infrared flux detection, which comes 
either from Infrared Astronomical Satellite (IRAS) or from Infrared
Space Observatory (ISO) data, are $\sim$ 50\% of the sources with
with a $\lambda L_{\lambda}(5100\dot{A})$ estimate. The PG QSOs of our 
sample are plotted as filled circles and all others as open circles. Both 
the optical and IR luminosities of the sources in our sample are very 
close to those of the mean of the IR-detected PG QSO population, 
indicating that our sources are representative of the most common PG 
QSOs but do not span the full range of AGN luminosities. In 
Fig.~\ref{fig:sample}, right panel, we plot the 60-25 \micron\ versus 
100-60 \micron\ color index $\alpha$ diagram (\citealt{degrijp}) of local 
PG QSOs; the symbols used are identical to those in the left panel of 
Fig.~\ref{fig:sample}. The MIR color indices of our sources are typical 
of all QSOs used in this diagram, again indicating that the sample we 
compiled is representative of the local IR-detected PG QSOs.

\subsection{Data acquisition and analysis}
\label{sec:reduction}

Our long-slit spectra were obtained with the ISAAC spectrometer 
(\citealt{moor98}) mounted on the Antu telescope unit of the ESO Very Large 
Telescope on Cerro Paranal, Chile. The QSO observations were performed in 
both service and visitor mode typically under excellent seeing conditions 
(optical seeing of $\sim$0\farcs5-0\farcs6) to minimize the effects of AGN 
light diluting the host signal. The integration time for each exposure was 
300 s to avoid saturation of the detector, and the total on-source 
integration time varied from 160 to 320 mins. The slit width was 
0\farcs6 and the detector scale is 0\farcs146 per pixel (see \pone).
The central wavelength was in the $H-$band and varied from 1.70 to 1.77 
\micron, depending on the redshift of each source and the 
wavebands of high atmospheric transmission. The spectral resolution 
$R=\lambda / \Delta \lambda$ was 5100. The slit position angles
(PAs) and the respective integration time per PA are tabulated in 
Table~\ref{tab:list} for the 12 sources observed. The redshift range of 
these QSOs ($0.050<z<0.112$) is on average lower than that of the ULIRGs
in our sample ($0.018<z<0.268$), and the integration times $\sim$2 times 
longer (on average), to ensure a high signal-to-noise ($S/N$) ratio of the 
host-galaxy signal. 

For most of the QSOs in this study, the $H-$band effective radius \reff\ 
and mean surface brightness $\mu_{\rm eff}$ within it are measured 
from Hubble Space Telescope (HST) NICMOS imaging data 
(\citealt{mcleod94a}; \citealt{veilleux06}). For some sources, 
NIR imaging data (assisted by adaptive optics) are also available from
\cite{guyon06}\footnote{It is not possible to use our acquisition images to 
derive \reff, as we did in Papers I and II for the ULIRGs, since the photon 
counts in our short-exposure-time acquisition images are strongly dominated 
by the point-like AGN, leading to \reff\ values biased towards small radial
extents.}. The quantities \reff\ and $\mu_{\rm eff}$ are given in 
Table~\ref{tab:list}. The \reff\ values are converted to the cosmology 
used in this paper ($H_0$=70~km~s$^{-1}$~Mpc$^{-1}$, $\Omega_{m}$=0.3, 
$\Omega_{\rm total}$=1).   

The data reduction tasks and procedures used to extract the host-galaxy 
spectra from the acquired ISAAC data are described in detail in Papers I 
and II. In the rest of this subsection, we 
mainly list the differences in the reduction method and the derived results 
between the ULIRG and the QSO data.

The main difference originates from the fact that photon counts associated with
the AGN continuum are roughly one order of magnitude greater than those of 
the host galaxy, leading to a significant suppression of the stellar absorption
features. For this reason, the highly nucleated AGN emission needs to be 
avoided for the accurate extraction of the stellar kinematics. To optimize 
the host $S/N$ ratio per slit, we extracted the spectra from the 
widest possible aperture (typically $\pm$1\farcs0-1\farcs4 from the 
center), excluding the very central region ($\pm$0\farcs3-0\farcs4). We 
then combined the results of the two slits into a single spectrum per object. 

The method of determining the stellar velocity dispersion $\sigma$\ from the 
reduced spectra was based on a Fourier correlation quotient technique 
(\citealt{bender}). This method uses a template spectrum to provide the 
Doppler-broadened profile of the stellar kinematics along the line-of sight 
(LOS). We fitted a combination of a Gaussian and a low-order polynomial to the 
broadened profile to determine the projected stellar velocity dispersion. 
The fit was performed to each bandhead individually and the quoted errors are 
the standard deviation of all measurements. In contrast to Papers I and II 
where $\sigma$ was measured from the central aperture, the stellar dispersion 
derived for the QSO hosts was typically measured at a radius of $\sim$\reff/4.
In simulations of gas-rich mergers, the velocity dispersion at the center 
and at the effective radius of the remnants tend to differ by $\sim$10\% 
(\citealt{beba00}). 
We ignored aperture
effects since the expected differences are typically within our error bars.
We could not extract information on the rotation of the host galaxy for the 
QSOs, given that it was necessary to perform pixel rebinning to achieve
the maximum possible $S/N$ ratio. 

\begin{figure*}
\centering
\includegraphics[width=8cm]{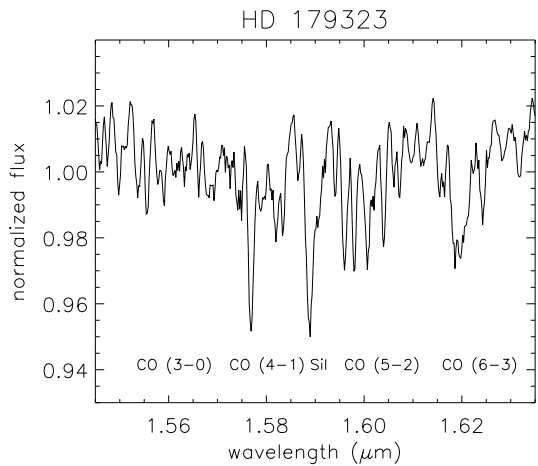}
\caption{The $H$-band spectrum of the K0Iab supergiant HD 179323 that
is used as a template for the extraction of the stellar kinematics of some 
QSO hosts.
\label{fig:K0} }
\end{figure*}

We observed four template stars for the extraction of the host velocity
dispersion: an M0III giant (HD 25472), an M1Iab supergiant (HD 99817),
a K5Ib supergiant (HD 200576), and a K0Iab supergiant (HD 179323). The spectra 
of the first three stars were presented in \cite{genzel01}. The K0I supergiant 
was observed with the integral-field-unit SINFONI on the VLT, which has a 
spectral resolution $R$=3000 in the $H$-band. 
The 2-d images were flat-fielded and converted into a wavelength-calibrated
3-d data cube using the SINFONI data cube reduction package ``SPRED'' 
(\citealt{spred}). After the extraction of the 1-d spectrum from the 3-d 
data cube, the stellar spectrum was sky-background subtracted and telluric 
features were accounted for by comparison with a stellar spectrum of known 
type. The methodology followed to apply these corrections to the 1-d spectrum
was identical to that applied for the ULIRGs in our program. The $H$-band 
spectrum of HD 179323 is presented in Fig.~\ref{fig:K0}. Unlike most ULIRGs,
which are best described by the M0III giant (see Papers I and II), some
PG QSOs are best fit by the K0I supergiant. In some cases, a linear 
combination of the two provides the best description of the QSO-host 
spectrum. The template used for each source and the measured velocity 
dispersion are given in Table~\ref{tab:masses}. The use of different stellar 
templates leads to $\sigma$ measurements that differ typically by 5 to 20\%. 

\begin{figure*}
\centering
\includegraphics[height=12cm,width=16cm]{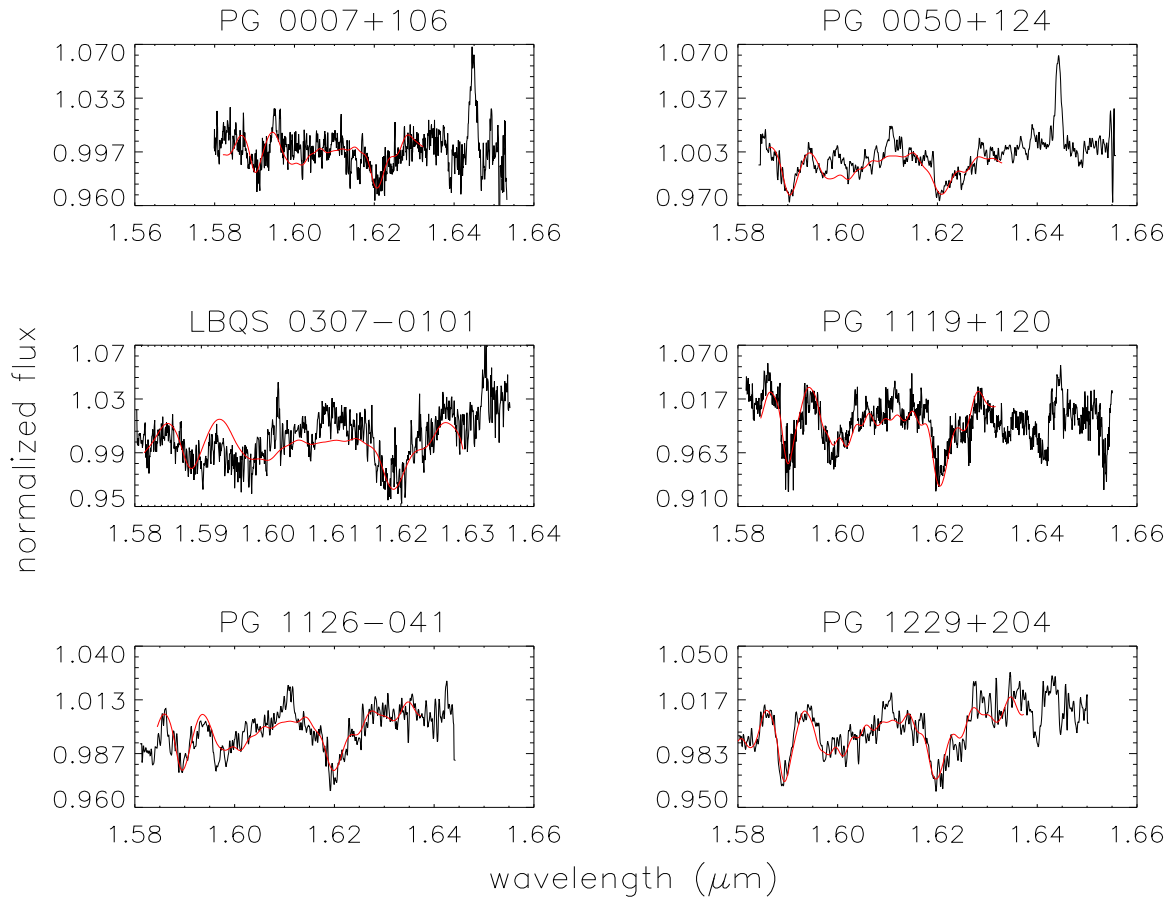}
\caption{The $H$-band spectra of the QSOs in this study. The selected 
stellar template for each source is overplotted as a solid line after
being convolved with the Gaussian that represents the LOS broadening 
function of the source. All spectra are shifted to the rest frame. The emission
lines occasionally seen at 1.611 \micron\ and 1.644 \micron\ correspond to 
Br 13 and [FeII] respectively. 
\label{fig:spectra} }
\end{figure*}

We plot the restframe spectrum of each QSO host galaxy in 
Fig.~\ref{fig:spectra}. For each source, we overplot the selected stellar 
template after convolving it with a Gaussian of dispersion equal to that 
measured. The spectra shown in Fig.~\ref{fig:spectra} are convolved to the 
resolution of the SINFONI observations for the sources that are best 
described by the K0I supergiant or the combination of the K0I and the 
M0III templates.

The velocity dispersion measurements can be affected by the strong dilution 
of the host-galaxy spectrum by the AGN continuum, which alters the shape of 
the wings of the stellar absorption features. To quantify this effect, we 
have constructed artificial QSO ISAAC data from which we measured the bulge 
velocity dispersion as we did for the real data. We convolved the spectrum 
of our M0III stellar template with a Gaussian of $\sigma=$200 \kms. We scaled 
its photon counts to those of a de Vaucouleurs bulge with the mean $H$-band 
magnitude and the mean redshift of the sources in our sample, $m_H$=13.5 mag
(\citealt{mcleod94a}; \citealt{veilleux06}) and $z=0.076$ respectively, that 
falls in a 0\farcs6 slit under seeing conditions of 0\farcs6. We diluted this
host-galaxy signal with a continuum whose strength was determined by counting 
the photons that originated from a point source of (mean) $m_H$=13.4 mag and 
that fell in the slit under the assumed seeing conditions. After adding 
Poissonian noise, we extracted the bulge dispersion in the same way we did 
for the real data. We repeated this procedure for 10 iterations and measured 
a mean $\sigma$ of 179 \kms\ with a standard deviation of 21 \kms. Therefore,
velocity dispersion underestimates of order 10\% due to the strong dilution 
of the host absorption features would not be unexpected. In addition to this, 
any differences between the stellar spectrum used to correct for the telluric 
features and the actual atmospheric absorption lines could also affect 
our spectra (see \pone). This could happen if, for example, the residuals from 
the removal of the stellar features are at $\sim$0.1\% continuum level, since 
the QSO-host-galaxy emission is only $\sim$ 10\% of the AGN continuum. Both 
these effects would lead to an artificial mismatch between the template and 
the galaxy spectrum, which is effectively included in the uncertainties of 
the measured velocity dispersion. In some cases, they lead to $\sigma$ error 
bars that are $\gtrsim$ 50 \kms.

\begin{figure*}
\centering
\includegraphics[width=11cm]{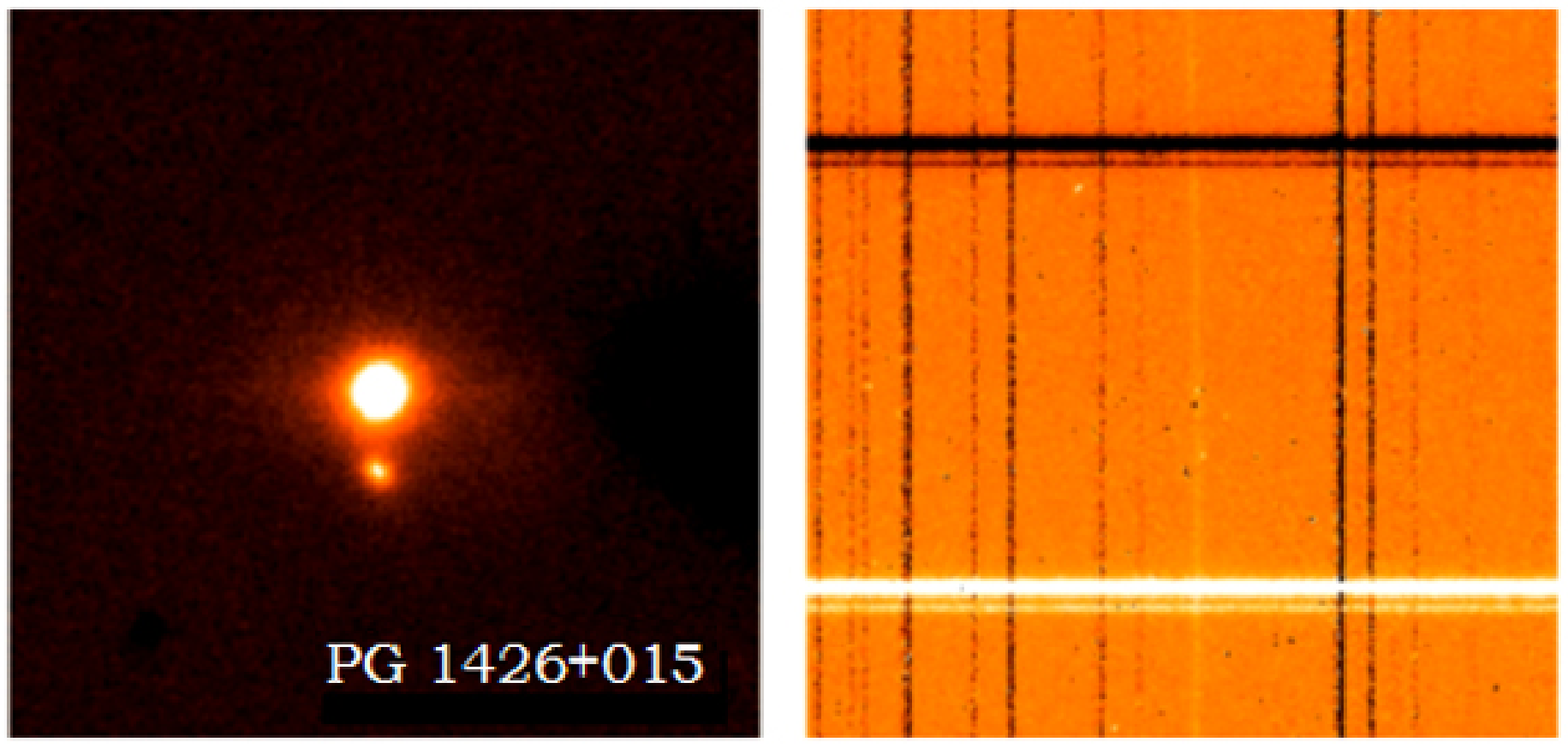}
\caption{ {\it Left panel:} The acquisition image of PG 1426+015 clearly 
shows that it is an interacting pair of 4.4 kpc nuclear separation. The
QSO corresponds to the NE nucleus. \newline
{\it Right panel:} The spectrum of PG 1426+015 for a slit passing through
both nuclei. The x and y axes have wavelength and spatial dimensions 
respectively. This image shows the difference between two 10-min integrations
at different detector positions; this technique is used to remove the sky 
lines. The number counts of the QSO spectrum are one order of magnitude 
greater than those of the faint nucleus. The velocity dispersions of the NE 
and the SW nucleus equal 185 ($\pm$ 67) and 154 ($\pm$ 27) \kms\ respectively.
\label{fig:pg1426} }
\end{figure*}

The $S/N$ ratio was insufficient for the detection of stellar absorption 
from the host galaxy of PG~1211+143, probably due to a very strong 
dilution of the host-galaxy light by the AGN continuum. PG~1426+015 is 
an interacting galaxy with two components. The data for both the bright,
north-east (NE) nucleus and the secondary, south-west (SW) nucleus are 
presented here (see Figs.~\ref{fig:spectra},~\ref{fig:pg1426}). The $H-$
band spectrum of the NE nucleus is mainly dominated by the AGN continuum 
and it is $\sim$30 times brighter than that of the fainter nucleus.


\section{Results}
\label{sec:results}
\subsection{Host-galaxy dynamical properties} 
\label{sec:host}

The stellar velocity dispersions of all QSO host galaxies in our sample are 
presented in Table~\ref{tab:masses}. The dispersion of the SW nucleus of PG 
1426+015 (which is not a QSO) is excluded from all statistical analyses of this
paper. The mean value of $\sigma$ is 186 \kms, with a standard deviation of 
$24$ \kms\ and a standard error (uncertainty of the mean) of 7 \kms.
Most of the sources of Table~\ref{tab:masses} with $\sigma <$ 200 \kms 
show spiral structure or tidal tails (PG~0050+124, PG~1119+120, PG~1126-041, 
PG~1229+204, PG1426+015, PG~2130+099) in their NIR images (\citealt{surace01}; 
\citealt{veilleux06}). 

For a stellar system characterized by a dispersion velocity $\sigma$, the
bulge mass is computed from 
\begin{equation}
\mbox{$m=1.40\times 10^6 \sigma^2 R_{{\rm eff}}$}, 
\end{equation}
where $\sigma$ is in units of \kms, \reff\ is in kpc and $m$ is in \msun\ 
(see \citealt{tacconi02}; \ptwo). Using this formula we find that the mean 
bulge mass of all QSOs in our sample is $2.09\times 10^{11}$ \msun. The bulge 
mass is somewhat lower than the dynamical mass, which also takes into account 
the stellar rotation. However, the fact that the mass scales with 
$3 \times \sigma^2$ and only \vrot$^2$ (\citealt{tacconi02}) and that 
any disk structure observed in the QSO hosts is not as prominent as the 
bulge (\citealt{veilleux06}) is a good indication that the dynamical mass 
will be similar to that of the bulge.
This result implies that the hosts of PG QSOs are typically $1.5 m_*$ galaxies,
for $m_*=1.4\times10^{11}$ \msun\ (see \citealt{genzel01}, \ptwo, and 
references therein). 

\begin{figure*}
\centering
\includegraphics[width=17cm]{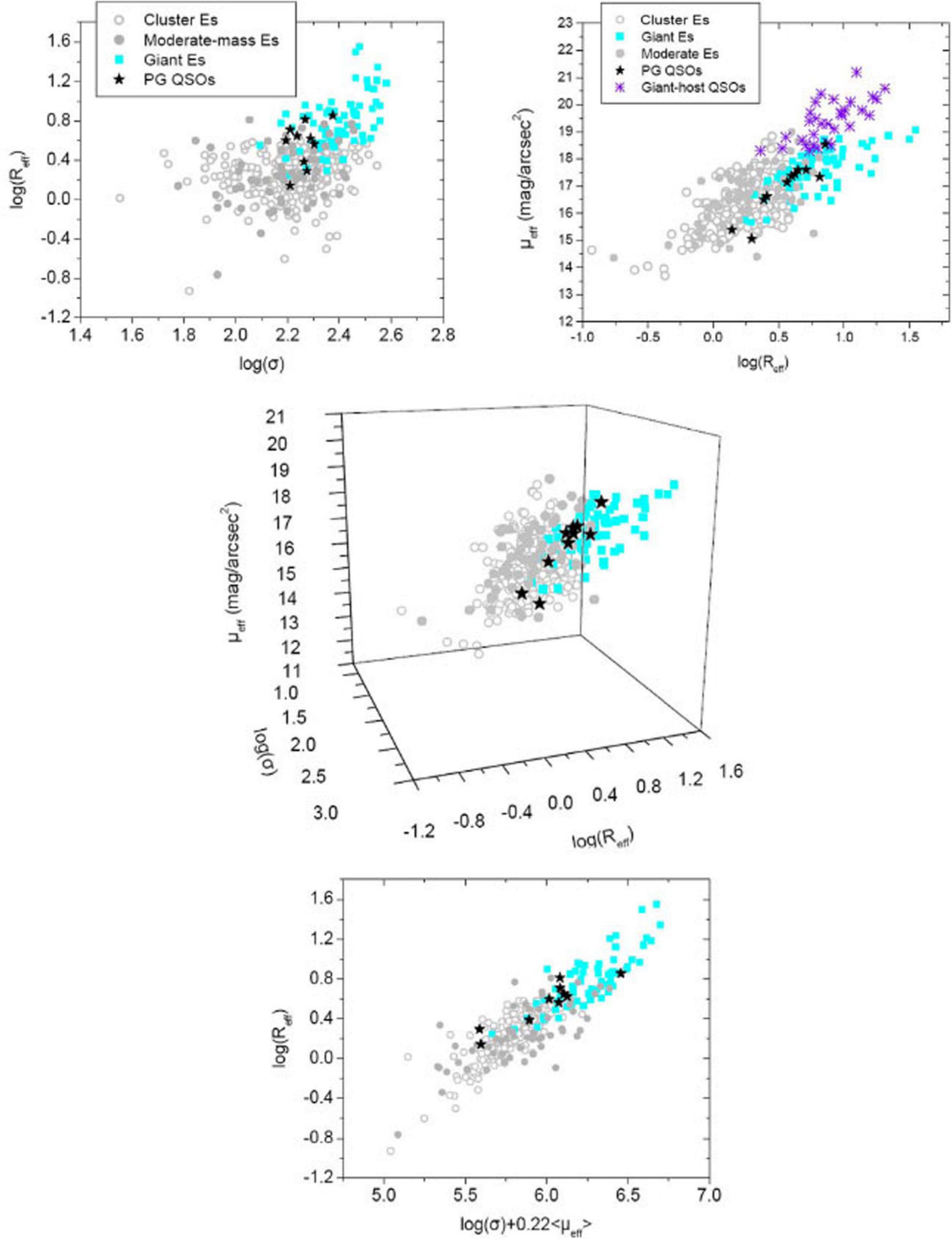}
\caption{The $K$-band fundamental plane of early-type galaxies. 
In all panels, giant boxy and moderate-mass disky Es (squares 
and circles respectively) are from \cite{bender92} and \cite{faber97}. More 
(cluster) Es (open circles) are from \cite{pahre}. The QSOs of this study 
are plotted as stars and those of \cite{dunlop03} as asterisks.
{\it Upper left panel:} The $\sigma$-\reff\ projection of the plane.
{\it Upper right panel:} The \reff -$\mu_{\rm eff}$ projection of the plane.
{\it Middle panel:} The 3-d view of the plane.
{\it Lower panel:} The plane projected as in \cite{pahre} for viewing clarity. 
\label{fig:fpe}}
\end{figure*}

In Fig.~\ref{fig:fpe} we place our QSOs on the $K$-band fundamental plane (FP)
of early-type galaxies (\citealt{djoda}; \citealt{dressler}) using the host 
effective radii and mean surface brightnesses from \cite{veilleux06} and 
\cite{mcleod94a} tabulated in Table~\ref{tab:list}. We converted $H$- into 
$K$- band host magnitudes using the Two Micron All Sky Survey (2MASS) 
mean $H-K$=0.3 color index correction (\citealt{jarrett})\footnote{We do not 
discriminate between elliptical or spiral hosts since the difference is 
small for the 2MASS galaxies and since the hosts of some of the QSOs in our
sample show patterns of spiral structure (e.g. \citealt{surace01}; 
\citealt{guyon06}).}. 
The data for the early-type galaxies are taken from \cite{bender92}, 
\cite{faber97}, and \cite{pahre} and converted to our cosmology. Giant boxy 
ellipticals (squares) are located on the upper-right part of the fundamental 
plane \reff -$\sigma$ projection, while moderate-mass ellipticals (circles) 
occupy the central and the lower-left parts (see Fig.~\ref{fig:fpe}, left 
panel). PG QSO hosts (stars) lie between moderate-mass and giant Es, but 
closer to the former, both in the $\sigma$-\reff\ projection that relates the
dynamical properties of the systems (upper left panel) and in the
3-dimensional view of the plane that takes into account their photometric 
properties (middle and lower panels). On the \reff -$\mu_{\rm eff}$ projection 
(upper right panel), also known as the Kormendy relation (\citealt{kormendy}), 
PG QSO hosts have only a small overlap with QSOs that are hosted by giant 
Es (\citealt{dunlop03}; see \S~\ref{subsec:compQSOs}).


\subsection{Black hole properties}
\label{sec:BH}

We compute the black hole masses of the QSOs in our sample by using the tight 
correlation between black hole mass and bulge velocity dispersion, the 
\msigma\ relation (\citealt{gebhardt00}; \citealt{feme00}). We use the 
\cite{tremaine02} formula,
\begin{equation}
\mbox{$M_{BH}=1.35 \times 10^{8} [\sigma/ 
200]^{4.02}$\msun,}
\end{equation}
as in Papers I and II. The application of the \msigma\ relation yields 
an average black hole mass of $1.12\times 10^8$ \msun\ (see 
Table~\ref{tab:masses}). 

While the \msigma\ relation was established using mostly local quiescent 
galaxies, it was subsequently shown that a similar relation exists in 
low-luminosity AGNs, namely type 1 Seyferts (\citealt{gebhardt00b}; 
\citealt{ferra01}; \citealt{nelson04}; \citealt{onken04}). The AGN black 
hole masses were measured from reverberation-mapping experiments 
(\citealt{blandford82}; \citealt{peterson93}; \citealt{peterson98};
\citealt{kaspi00}; \citealt{peterson04}; \citealt{kaspi05}; \citealt{veste06}).
Reverberation mapping measures the size of the broad-line region (BLR) from 
the light-travel time delay between continuum and emission-line flux 
variations. Under the assumption that the BLR gas is virialized, the central 
black hole mass is given by
\begin{equation}
\mbox{\mbh$ = f R_{\rm BLR} \Delta V^2/G$,} 
\label{eq:vp}
\end{equation}
where $R_{\rm BLR}$ is the size of the BLR as measured by the time delay, 
$\Delta V$ is the emission-line width, $G$ is the gravitational constant, and 
$f$ is a dimensionless scaling factor that encapsulates uncertainties in the 
structure and inclination of the BLR. The mean value $\langle f \rangle = 5.5$ 
(\citealt{onken04}), is statistically determined by assuming that the \msigma\ 
relation is the same in both active and quiescent galaxies. According to 
\cite{veste06}, use of this single value for $f$ leaves residual scatter 
around the \msigma\ relation that indicates that the reverberation-based 
black hole masses are accurate to a factor of $\sim$ 3. However, in individual 
cases the actual value of $f$ depends on currently unknown factors, such as 
the inclination of the BLR (\citealt{collin06}). 

\begin{figure*}
\centering
\includegraphics[width=8.6cm]{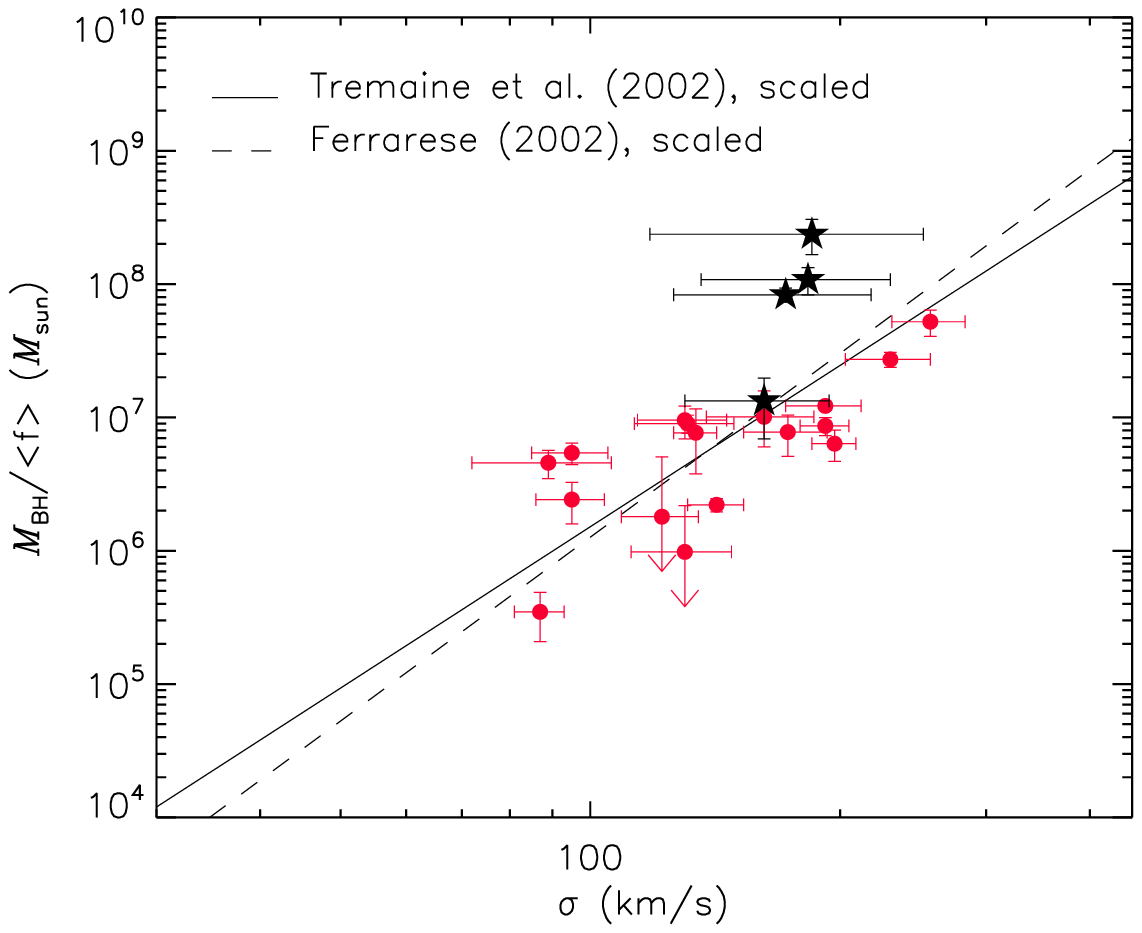}
\caption{ 
The \msigma\ relation for AGNs with direct \mbh\ measurements. In this 
Figure, we plot the virial product \mbh/$\langle f \rangle$ of reverberation 
experiments (\citealt{peterson04}) versus the host velocity dispersion
to avoid uncertainties in the mean value of the factor f. The Seyfert 1 AGNs 
of \cite{onken04} and \cite{nelson04} are shown as circles. A few 
low-luminosity AGNs with upper limits on their \mbh\ values are indicated
by arrows pointing down. The QSO datapoints are shown as stars. The solid and 
the dashed lines correspond to straight lines with slopes identical to those 
of the quiescent-galaxy relations (\citealt{tremaine02} and \citealt{ferra02} 
respectively) that fit the AGN datapoints (\citealt{onken04}). 
\label{fig:msigma}}
\end{figure*}

Reverberation BH masses are available in the literature (\citealt{peterson04})
for four of the PG QSOs in our sample that have a host galaxy velocity 
dispersion measurement (see Table~\ref{tab:masses}). We place these sources 
on the AGN \msigma\ relation in Fig.~\ref{fig:msigma}, where the QSOs 
(indicated by stars) are plotted over the AGNs (circles) of \cite{onken04} and 
\cite{nelson04}\footnote{For the sources with two $\sigma$ measurements we use 
the average of the two.}. To avoid uncertainties in the mean value of $f$,
we use the virial product \mbh/$\langle f \rangle$ of Eq.~\ref{eq:vp}. The 
solid and dashed lines correspond respectively to the \cite{tremaine02} and 
\cite{ferra02} fits, scaled down by a factor of $\langle f \rangle$ to match 
the AGN datapoints (\citealt{onken04}). We find that one of our sources, 
PG~1229+204, falls on the AGN relation, showing good agreement between the 
dynamically-determined and the reverberation-based black hole mass 
measurements. However, three of the four reverberation-mapped QSOs lie 
above the locus of Seyfert 1 AGNs with similar velocity dispersions. Thus, 
the average dynamically-determined black-hole 
masses for PG QSOs tend to be smaller than those measured by reverberation 
experiments. This result might be attributable to incorrect assumptions, 
measurement errors, or simply small number statistics. We consider each of 
these in turn.
 
By incorrect assumptions, we are referring principally to the underlying 
assumption that a single, statistically determined value of $f$ is equally
good for all AGNs. The apparent differences can also reflect real differences 
in the host and black-hole properties between lower-mass and higher-mass 
black holes, such as those claimed for quiescent galaxies by \cite{lauer}. 
Alternatively, the offset
may be indicating differences due to inclination effects (\citealt{collin06})
or to a currently misestimated calibration of the factor $f$. This 
possibility is implied by Fig.~\ref{fig:msigma}: the virial products of the 
PG QSOs in our sample lead to an offset above the AGN \msigma\ relation
that brings the sources on the relation for quiescent galaxies without the use 
of a factor $f$ (in other words, with $\langle f \rangle \sim 1$). Our result 
implies that the mean factor $f$ could be smaller than 5.5. 

PG QSOs could also deviate from the AGN \msigma\ relation due to 
measurement errors either in their stellar velocity dispersions or in their 
\mbh\ estimates. We suspect that PG~2130+099 represents a case of 
the latter. The reverberation-based mass significantly exceeds the mass 
predicted by \cite{veste06} for the optical luminosity and {\rm $H\beta$} 
flux of this source (see below). The object is also an outlier in the BLR 
radius $-$ luminosity relation (\citealt{kaspi00}; \citealt{bentz});
the BLR size, based on reverberation measurements, appears to be much too 
large for its luminosity. An overestimated value of the BLR radius is 
plausible for this object since its light curve was not particularly well 
sampled (\citealt{kaspi00}).

It is also possible that there are systematic errors in the measurement 
of $\sigma$, e.g., from dilution of the stellar light by the AGN continuum. 
However, such differences can typically be of order $\sim$10\% (see 
\S~\ref{sec:reduction}). By themselves, they seem unable to account for the 
discrepancy between the dynamically-determined and the reverberation \mbh\ 
estimates, which on average is a factor of 7. However, other systematics, such 
as stellar population effects might also contribute to the discrepancy. Errors 
of this type could be introduced by the use of NIR velocity dispersion 
measurements for the construction of the AGN \msigma\ relation, which is based 
on optical $\sigma$ values. Discrepancies of order 10\% have been reported in 
the literature between CO and various optical dispersion measurements, mostly
for quiescent galaxies that contain disky structures (e.g., \citealt{silge}; 
\citealt{oliva99}). However, the corrections implied from the literature
are not only small, but also have conflicting signs. It is very likely that 
they reflect differences in the methodology, spectral resolution, and stellar 
templates that have been used by the various authors. At this point, the 
possibility that there are true systematic differences between $\sigma$ 
measurements in the optical and NIR cannot be discounted. Further data are 
required to either confirm or quantify such effects. 

\begin{figure*}
\centering
\includegraphics[width=8.6cm]{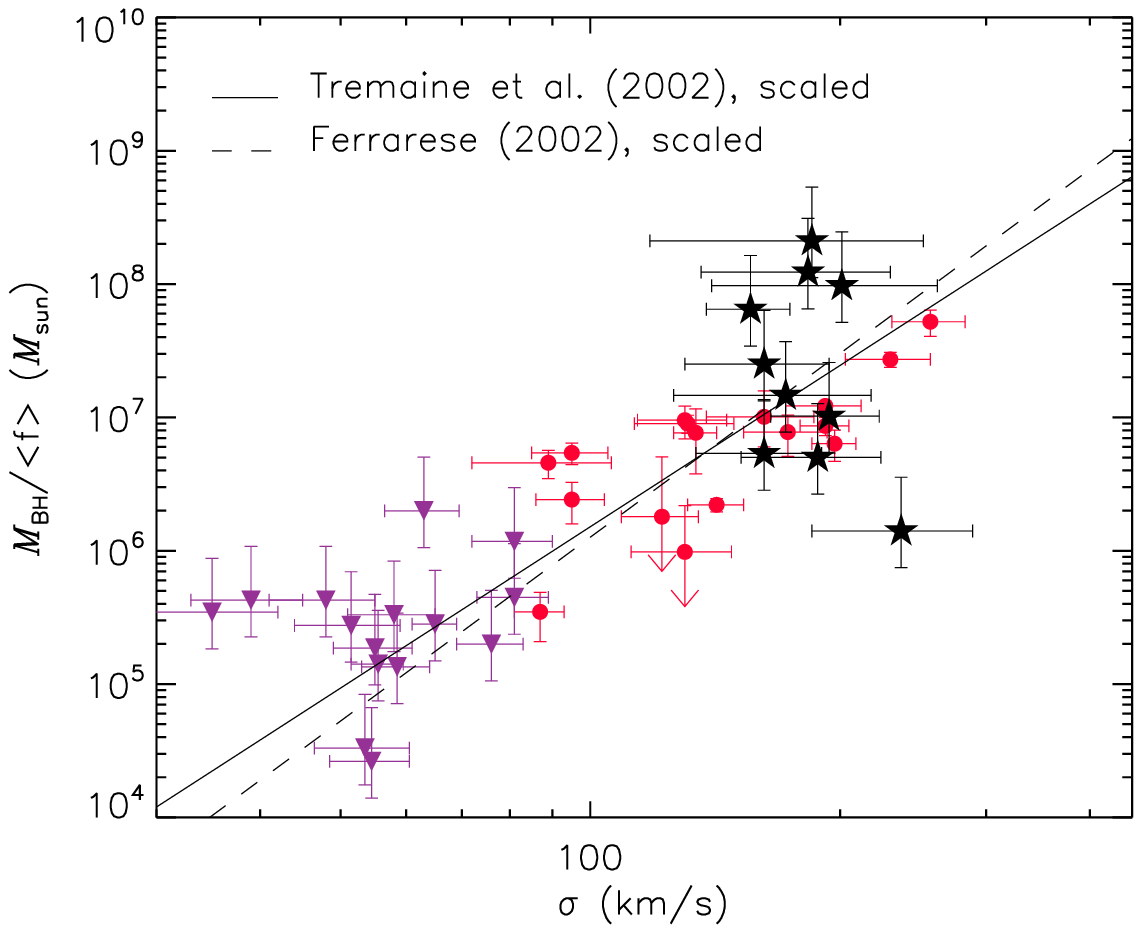}
\caption{ 
The AGN \msigma\ relation studied at its high- and low- mass end using 
indirect, single-epoch, \mbh\ estimates for local PG QSOs (\citealt{veste06}) 
and low-mass AGNs (of \mbh\ $\sim 10^6$; \citealt{barth}). As in 
Fig.~\ref{fig:msigma}, the value $\langle f \rangle=5.5$ is divided out of 
all black-hole estimates. The symbols used for the Seyfert 1 AGNs of 
\cite{onken04} and \cite{nelson04} and the PG QSOs of this study are identical 
to those used in Fig.~\ref{fig:msigma}. The low-mass AGNs are plotted as 
triangles. The solid and the dashed lines correspond to the \cite{tremaine02} 
and \cite{ferra02} fits scaled to the AGN datapoints.
\label{fig:msigma_in}}
\end{figure*}

A third possible explanation is that the apparent displacement of the PG QSOs 
from the AGN \msigma\ relation is simply an artifact of small-number 
statistics. As a test of this possibility, we use a secondary method to 
estimate the 
BH masses for 10 objects in our sample. Reverberation-mapping results show 
a strong correlation between BLR radius and the luminosity of the AGN (Kaspi 
et al.\ 2000; 2005; Bentz et al.\ 2006); it is therefore possible to infer 
the BLR radius and the black-hole mass of the AGN by making use of an 
optical (or UV) flux measurement and a ``single-epoch'' spectrum that 
provides emission-line widths (e.g., \citealt{veste02}). Using such relations, 
\cite{veste06} have recently provided \mbh\ estimates for PG QSOs without 
direct reverberation measurements. Of those, we use the values computed
from the width of the $\rm H\beta$ line and the continuum luminosity at 
5100\,\AA. The statistical error bars\footnote{
\cite{veste06} provided statistical uncertainties for this relation at various
confidence levels for the sample as a whole and measurement-related error bars
for individual sources. The statistical uncertainties are significantly
higher than those of the flux measurements, leading to actual errors in
the individual \mbh\ estimates that can be as high as an order of magnitude
(\citealt{kelly}).}
that we ascribe to these black-hole mass estimates are based on the 1-sigma
scatter around this luminosity-black hole relation, which amounts to
0.52 dex or a factor of 3.3 (\citealt{veste06}). In Fig.~\ref{fig:msigma_in}, 
we plot the 10 PG QSOs (shown as stars) with single-epoch \mbh\ estimates over 
the Seyfert 1 AGNs of \cite{onken04} and \cite{nelson04} (circles). We also 
overplot the \mbh $\sim 10^6$ \msun\ AGNs of \cite{barth} (triangles) to 
simultaneously display the behavior of the \msigma\ relation at the 
high and low-mass ends. For all sources, we use the $\mbh / \langle f 
\rangle $ estimates with $ \langle f \rangle $ = 5.5, as in 
Fig.~\ref{fig:msigma}. By increasing the number of sources on the AGN 
\msigma\ diagram, we find a better agreement between the Seyfert 1 AGNs and 
the PG QSOs. The scatter in the relation seems to be larger at both the 
low and high mass ends than for the reverberation-mapped AGNs, due to the 
additional scatter that is intrinsic to the single-epoch scaling relations. 

On the basis of Fig.~\ref{fig:msigma_in}, it appears that the result found 
for the reverberation-mapped PG QSOs is at least partially ascribable to 
small number statistics, and that any correction to $\langle f \rangle$ is 
rather small. Moreover, the offsets between the single-epoch and the 
dynamically-determined values of \mbh\ are comparable both at the high and 
low mass ends of the \msigma\ relation; specifically, 
$< \delta \log$[\mbh]$>$ is 0.15 and $\sim$0.25 for the PG QSOs and the 
low-luminosity \cite{barth} samples, respectively. That the offset is of
the same sign in both ends of the relation can be traceable to the fact that 
some of the QSOs in our sample and those of \cite{barth} are narrow-line 
Seyfert 1 (NLS1) galaxies, whose black holes may be related in a different 
way to the host galaxy (\citealt{grupe}; \citealt{barth}). That the offset 
between these two samples and that of the \cite{onken04} AGNs is small is an 
indication that the relation that connects the bulge dispersion to the 
black-hole mass is global, applying to AGNs with \mbh\ from $\sim$10$^5$ to 
$\sim 5 \times$ 10$^8$ \msun. However, the measurement 
of $\sigma$ in larger samples of reverberation-mapped AGNs is a crucial 
step towards understanding where differences may come from and presenting 
a new best fit to the AGN data.

To estimate at what rates our sources accrete, we calculate their Eddington
black hole masses, $M_{BH}({\rm Edd})$, from their Eddington luminosities, 
$L_{{\rm Edd}}$, as 
\begin{equation}
\hbox{$L_{{\rm Edd}} / L_{\odot}=3.8 \times 10^{4} M_{\rm BH}({\rm Edd})/  M_{\odot}$.}
\end{equation}
We consider that all of the QSO bolometric luminosity comes from the 
active nucleus (in contrast to \pone\ and \ptwo\ where we assigned half of 
the infrared luminosity of ULIRGs to $L_{{\rm Edd}}$). We compute the 
bolometric luminosity from the optical luminosity (see Table~\ref{tab:list}) as
\begin{equation}
\hbox{$L_{bol}=C*\lambda L_{\lambda}(5100\dot{A})$,}
\end{equation}
where the conversion factor $C$ is $\sim9$ 
(\citealt{kaspi00}). The value of $C$ can significantly differ from one 
QSO to another; therefore, the accuracy of its mean is rather small (within 
a factor of 3; \citealt{sanders89}; \citealt{elvis94}; \citealt{netzer03}). 
The estimated Eddington accretion efficiencies, 
$\eta_{\rm Edd} \equiv M_{\rm BH}({\rm Edd})/ M_{\rm BH}$, 
carry this uncertainty. Furthermore, since part of the emission 
originates from the host, the values of $\eta_{\rm Edd}$ given in 
Table~\ref{tab:masses} should be treated as upper limits. The mean 
Eddington accretion efficiency of the QSOs in our sample is 0.25, in 
good agreement with \cite{mcleod99}.


\section{ULIRG vs. (IR-excess) PG QSO dynamics}
\label{sec:selection}

To properly investigate whether QSOs and ULIRGs may be related, it 
is of particular importance to identify those QSOs with IR-excess 
emission and compare the dynamics of the latter to those of ULIRGs. Several 
criteria have been used in the literature to identify IR-excess QSOs (see 
\citealt{sanders89}, \citealt{surace01}, \citealt{lipari}, 
\citealt{canalizo01}, and references therein). Some were based on 
the 60 \micron\ flux and the 25/60 \micron\ flux ratio. \cite{sanders89} 
and \cite{surace01} defined IR-excess QSOs to be those sources with 
\lir/$L_{\rm bb} >0.46$. \cite{lipari} and \cite{canalizo01}, among others,
used the diagram of MIR color indices (Fig.~\ref{fig:sample}, right panel)
to select sources in a possible transition phase between starburst and QSO 
activity. The classification is based on the fact that AGN-dominated (power 
law) sources occupy a different locus than starbursts, whose characteristic 
interstellar dust temperature is $10-10^{2}K$ (\citealt{dowso98}). We opt to 
use the definitions that are based on all IRAS bands, since they are less 
susceptible to the thresholds set, which are sometimes subjective and specific
to the sample under examination. We refine the IR-excess definition to take 
into account both the ratio of the IR over the optical/ultraviolet luminosity 
and the MIR color-index diagram. As IR-excess we classify those sources that 
have \lir/$L_{\rm bb} >0.46$  and lie between the AGN- and starburst- 
dominated loci on the diagram of MIR color indices.

Five sources satisfy our set of IR-excess criteria, namely, PG 0050+124, 
PG 1119+120, PG 1126-041, PG 1426+015, and PG2130+099. 
The mean velocity dispersion of this sample is 180 \kms\ with a standard 
deviation of 13 \kms\ and error of 6 \kms. Statistically, the difference
between this value of $\sigma$ and the mean of our entire sample is smaller
than 1 Gaussian sigma, so we find no particular reason to separately study
the dynamical properties of IR-excess QSOs and all PG QSOs in our sample. 
The comparison that follows refers to our sample as a whole.

The mean velocity dispersion of the QSO hosts is 1.16 times that of the ULIRG 
remnants in our sample, 161 \kms (\ptwo). This difference 
in the dispersion can account for differences in the masses of ULIRGs and 
QSO hosts of the order $\sim 5\times10^{10}$ \msun. However, the observed 
difference is $\sim 10^{11}$ \msun, since the mass of the ULIRG remnants 
is 8.91 $\times10^{10}$ \msun\ (\ptwo) and that of the QSO hosts is 
2.09 $\times10^{11}$ \msun. That the measured difference is that large 
can mainly be attributed to the effective radii of QSOs which are roughly
twice as large as those of ULIRGs; the mean value of \reff\ is 3.9 kpc 
and 2.2 kpc for the QSOs and the ULIRG remnants in our sample 
respectively\footnote{ According to \cite{tacconi02}, the measured effective 
radii of the gas-rich merger remnants may increase when the ultraluminous 
phases cease, since the strongly nucleated starburst emission will eventually 
fade away. Under this hypothesis, the difference in the stellar mass of the 
QSO hosts and the elliptical galaxies that ULIRGs will finally form may be 
smaller than that measured at present time.}.

\begin{figure*}
\centering
\includegraphics[width=17.0cm]{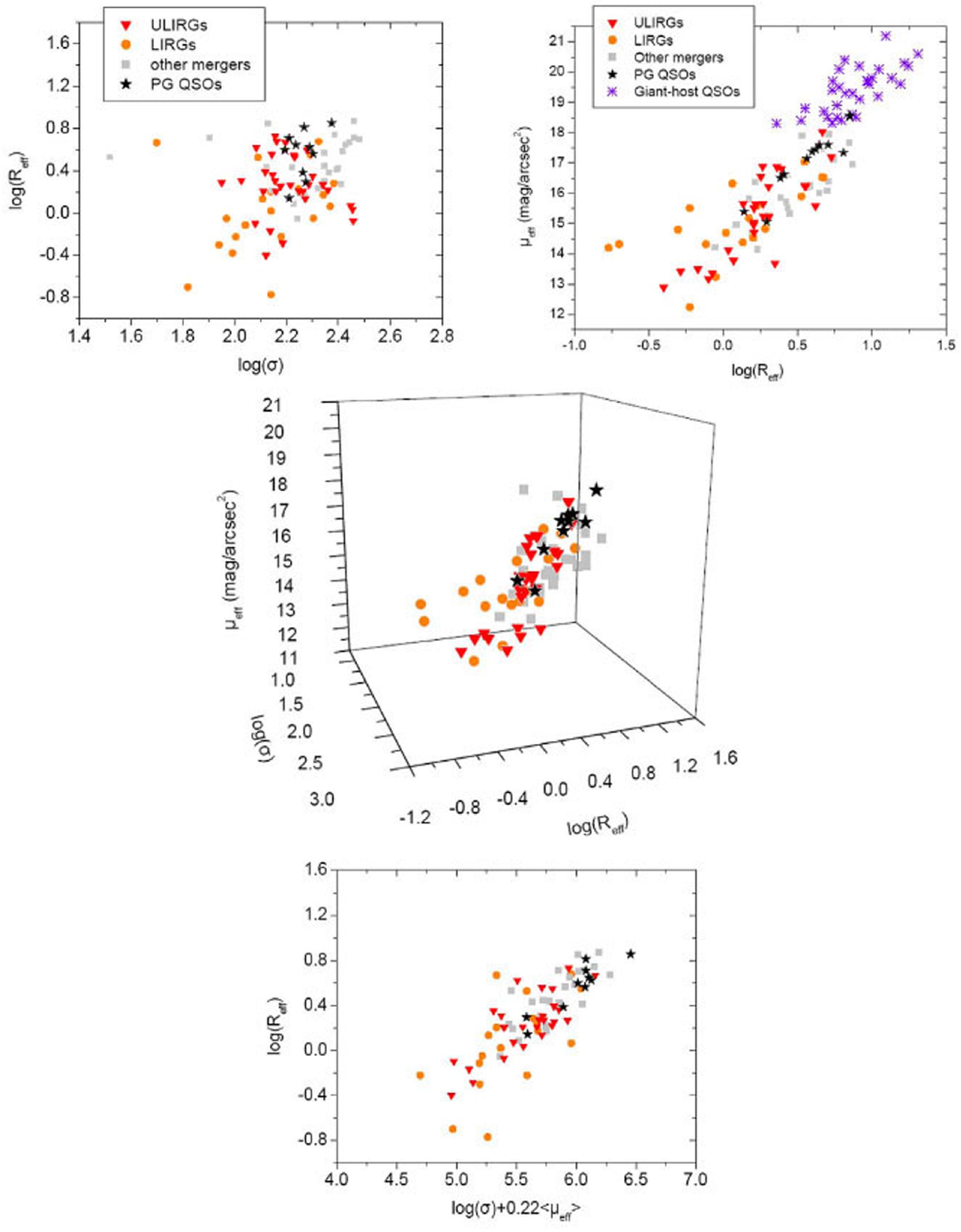}
\caption{The $K$-band fundamental plane of merger remnants. 
The panels are identical to those of Fig.~\ref{fig:fpe}. 
In all panels, ULIRG, LIRG, and other (visually-selected)
remnants are plotted as triangles, circles, and squares respectively.
The merger remnant data are from \cite{genzel01}, \cite{tacconi02}, \ptwo, 
\cite{rothberg}, \cite{hinz}, \cite{shier}, and \cite{james}. The QSOs of 
this study are plotted as stars and those of \cite{dunlop03} as asterisks.
\label{fig:fpm}}
\end{figure*}

We place both PG QSOs and ULIRGs on the fundamental plane of early-type 
galaxies (Fig.~\ref{fig:fpm}), to illustrate the similarities and differences 
amongst their galaxy properties. To better visualize the region that each 
population occupies, we construct various 2-d and 3-d views of the plane, as
in Fig.~\ref{fig:fpe}. Local ULIRGs (\citealt{genzel01}; \citealt{tacconi02}; 
\ptwo; \citealt{rothberg}) are plotted as triangles and PG QSOs as stars.
For comparison, we also use local LIRGs (\citealt{shier}; \citealt{james}; 
\citealt{rothberg}; \citealt{hinz}), which are plotted as circles. Other,
morphologically-selected merger remnants that are not IR-bright
(\citealt{rothberg}) are plotted as boxes. The latter category can include 
mergers of various types (e.g. spiral-spiral, elliptical-spiral, 
elliptical-elliptical); it can also 
include former ULIRGs and LIRGs whose starbursts have now faded away. For 
viewing clarity we do not overlay any early-type galaxies on this figure. 
The position of PG QSOs on the dynamical $\sigma$-\reff\ projection shows 
that they lie at the upper end of the locus of ULIRGs and LIRGs and in the 
middle of the locus of morphologically-selected mergers; therefore a subset 
of all (U)LIRGs and PG QSOs have identical dynamical properties.


\section{Discussion}
\label{sec:comp}

\subsection{Relation of PG QSOs to gas-rich mergers}
\label{subsec:compmergers}

In this section, we place our results on the host kinematics of PG QSOs into a 
wider evolutionary framework by combining multiwavelength spectroscopic and 
imaging evidence that links PG QSOs to mergers or interactions of galaxies 
that possess gas. Some sources  show clear signs of interaction, such as 
disturbed morphology (e.g. PG~1211+143; PG~2214+139; \citealt{surace01}; 
Guyon et al. 2006) and tidal tails (e.g. PG~0007+106; \citealt{surace01}).
Faint blobs, which could be identified as secondary nuclei, appear in the 
images of PG~2130+099 and PG~0007+106 (\citealt{veilleux06}; \citealt{guyon06}.
Various imaging analyses indicate that several sources in our sample 
are unambiguously binary systems. \cite{surace01} found an elongated companion
north-west of the main nucleus of PG~1119+120. According to \cite{veilleux06} 
and \cite{guyon06}, PG 1126-041 also interacts with a galaxy at a projected 
distance of 6.6 kpc. \cite{stockton82} and \cite{canalizo01} spectroscopically 
confirmed that PG~0050+124 has a companion galaxy 16.5 kpc to the west of the 
QSO. Hubble Space Telescope (HST) imaging data presented by \cite{schade} 
indicate the presence of a second nucleus south-west of the bright nucleus 
of PG~1426+015 (also see Fig.~\ref{fig:pg1426}). Our absorption-line 
spectroscopy of the host of PG~1426+015 confirms that the redshift of the 
second nucleus is identical to that of the QSO. The projected nuclear 
separation of the interacting galaxies is 4.4 kpc. PG~1426+015, PG 1126-041, 
and PG~0050+124 are amongst the objects in our sample for which the QSO 
phase has been reached already before the individual nuclei 
coalesce\footnote{On the other hand, PG~1119+120 does not officially fulfill 
the QSO definition, since its bolometric luminosity is $<$ 10$^{12}$ \lsun\ 
(\citealt{surace01}).} (see also \citealt{mcleod94b}). Therefore, QSO phases
are not necessarily linked to the end of the merger or interaction process.

Various pieces of evidence for star-formation exist in several of the 
PG QSOs in our sample.
\cite{canalizo01} have modeled the optical spectra of 9 ULIRGs and QSOs
suspected to be in a possible ULIRG/QSO transition phase (from their MIR
color-index diagram). They find that the spectra of most of their objects 
are described by a combination of old and recently formed stellar
populations. The ages derived for the starbursts were $<300~Myr$. According 
to \cite{canalizo01}, one of the sources in our sample, PG~0050+124 
(or I~Zw~1) has ongoing star formation (\citealt{schinnerer98}). 
In the MIR, {\it Spitzer} IRS spectroscopy of QSOs and
ULIRGs in our QUEST project (PI Veilleux) indicates the presence of
starbursts in PG QSOs. Polycyclic Aromatic Hydrocarbon (PAH) molecule 
emission has been individually detected for 11 of the 26 QSOs observed; 
for the remaining sources, stacking of the spectra also reveals the 
presence of PAHs (\citealt{schweitzer06}). Based on the strength of
the PAH emission, \cite{schweitzer06} find that star formation in PG QSOs 
is much stronger than what indicated by emission lines at shorter wavelengths, 
such as the [OII] line at 3727 $\dot {\rm A}$ (\citealt{ho05}), which can 
be affected by obscuration.

The presence of gas, which is required to initiate star-formation and 
AGN-accretion events, has been confirmed in several PG QSOs. Molecular
CO and HCN J=(1-0) emission lines have been detected by Evans et al. (2001; 
2006), and \cite{sco03} in a total sample of 18 PG QSOs. The CO luminosities 
were at most an order of magnitude lower than those of ULIRGs 
(\citealt{evans01}; 06; \citealt{dowso98} and references therein). 
Converting the CO luminosity into gas mass implied that PG QSOs can have 
molecular gas reservoirs as massive as $\sim$10$^{10}$ \msun. For this
computation, a Galactic CO-to-H$_2$ conversion factor of 4 \msun $K^{-1} 
[km/s]^{-1} pc^{-2}$ was used. However, this gas mass may be an upper limit, 
if the conversion factor for PG QSOs is closer to that of ULIRGs, 0.8 
\msun $K^{-1} [km/s]^{-1} pc^{-2}$ (\citealt{dowso98}), than to that of 
the Milky Way, due to a potentially compact, high-pressure gas distribution 
in the QSOs. 

In some cases, the evidence that relates local PG QSOs and ULIRGs is 
strong. For example, a few sources are known to fulfill both 
the ULIRG and QSO classification criteria (e.g. PG~0157+001; PG~1226+023).
More often, the IR luminosity output is similar to that of 
LIRGs. This result is still consistent with an evolutionary scenario 
suggesting that one population evolves into another, since some LIRGs are 
plausibly the progenitors or descendants of ULIRGs. However, the LIRG 
population as a whole is believed to originate from a wider variety of 
interactions or mergers, e.g., from mergers of larger progenitor mass ratios 
than those of ULIRGs (Ishida et al. 2007, in preparation). This is an 
argument against a scenario where all ULIRGs undergo a QSO phase (and 
vice-versa), and in favor of a scenario where QSO activity is triggered 
by progenitors or initial conditions that are similar 
but not identical to those of ULIRGs. Examples include some PG QSOs in
our sample (PG~1126-041, PG~1229+204, and PG~2130+099) that are amongst 
the sources with the lowest optical luminosity and that have signatures of 
spiral structure (\citealt{surace01}; \citealt{veilleux06}). If these sources 
have indeed gone through a recent interaction, they may have been involved in 
unequal-mass major ($\sim$3:1) or minor ($>$3:1) mergers, since the latter 
are known to form remnants with significant angular momentum (e.g., 
\citealt{beba00}). Therefore, it seems that both scenarios about 
the relation of ULIRGs and QSOs are partially valid.

\subsection{Relation of PG QSOs to other QSO populations}
\label{subsec:compQSOs}

It is important to point out that local gas-rich mergers are not the direct 
progenitors of all local QSOs, and that the PG QSO sample does not include 
objects 
representative of all local QSO populations (\citealt{veste06}). Selection 
effects and differences between the Palomar BQS and the Sloan Digital Sky
Survey (SDSS) samples have 
been extensively studied by \cite{jester05}. The completeness of the BQS is 
reported to be smaller than that of other surveys such as the SDSS and the 
Hamburg-ESO Quasar Survey (\citealt{lutzwi}). Due to its $B$-band magnitude 
cutoff, $m_B \lesssim 16$, the BQS sample selects bright local AGNs, which 
are likely to be presently accreting at high rates (see \S~\ref{sec:sample}). 

A sample compiled in an opposite manner, 
i.e. with magnitudes fainter than $\sim$16 mags, will probably include AGNs 
that accrete at lower rates than PG QSOs. Such a sample is the $V$-band
magnitude $m_V \gtrsim 15$ radio-loud (RL) or radio-quiet (RQ) QSO
population at $0.138<z<0.258$ of \cite{dunlop03}. These authors selected 
objects fainter than 15 mags in the $V$-band, to match the optical properties
of their entire sample to those of their RL subsample (\citealt{dunlop93}). 
These RL QSOs and their optical RQ counterparts have, indeed, supermassive 
black holes of 10$^8$-10$^9$ \msun\ that currently accrete at low rates 
($\sim$0.055 of the Eddington value) and are typically located in 
massive elliptical hosts (of mean bulge mass $5.6\times10^{11}$ \msun). 
\cite{tacconi02}, \cite{veilleux02}, and \cite{dunlop03} showed that ULIRGs 
and giant-host QSOs have a very small overlap in their host photometric
properties and extents, which indicates that their origin may be sought
in different mechanisms. We find that the mean dynamical properties of PG 
QSOs are intermediate to those of these two populations, but significantly 
closer to those of ULIRGs.

A similar conclusion is derived from the comparison to SDSS radio-loud
sources. \cite{best05} have found a correlation between radio-activity and 
the size of the host galaxy and black hole mass; the bulk of radio emission 
mainly originates from the sources with the most massive black holes 
($\log[\mbh]\gtrsim 8.2$) and the most massive bulges in their sample. The
latter seem to have host-galaxy properties closer to those of the 
\cite{dunlop03} sample than those of the PG QSOs in this study.

A possible scenario for the local formation of $5 \times 10^8$-10$^9$ 
\msun\ BH QSOs is by the reignition of a 10$^7$-10$^8$\msun\ black hole 
by the delivery of new gas due to tidal interactions. In the case of
local galactic mergers this delivery could happen, for example, via 
elliptical-spiral or elliptical-elliptical mergers, implying that a
sequence of merging events would be required for the build-up of the 
black hole. A more likely possibility is that these QSOs formed at earlier 
epochs, similarly to some giant Es (see \ptwo\ and references therein).


\section{Conclusions}
\label{sec:conc}
We have acquired $H$-band long-slit spectra of 12 QSOs, mainly drawn from the
PG catalog, to study the dynamical properties of their hosts and investigate 
whether they have an origin analogous to that of ULIRGs. The compiled sample 
is representative of local IR-detected PG QSOs and appropriate for 
investigating the triggering of QSO activity by mergers. We find that:

\begin{enumerate}
\item
The long-integration, excellent-seeing-con\-di\-tion spectroscopic data 
obtained from the VLT have enabled us to directly extract host-galaxy 
velocity dispersion and dynamic properties of local QSOs in the $H$ band. 
The individual stellar velocity dispersions vary from 156 to 237 \kms. The 
mean $\sigma$ value of the sample is 186 \kms\ with a standard deviation
and error of 24 and 7 \kms. The host galaxies have bulges of $\sim$1.5
$m_*$ stellar mass. The dynamically-determined black-hole masses of the 
PG QSOs are of order $\sim 10^8$\msun, and they accrete at Eddington 
efficiencies of order $\sim 10^{-1}$.  
\item
The measurement of the host velocity dispersion enables us to place 4 
reverberation-mapped PG QSOs on the AGN \msigma\ relation, acknowledging 
the large error bars and possible systematics of the measurement. Most are 
located above the AGN best-fit functions. When using indirect, single-epoch, 
\mbh\ measurements for 10 QSOs, the mean offset from the best fit to the 
AGN data is rather small. The high- and low- mass ends of the relation 
seem consistent with each other, implying that this relation probably 
applies to AGNs with a variety of \mbh\ sizes.  
\item
The differences in the host-galaxy and black-hole properties of our entire 
sample and the IR-excess subsample are insignificant. The position of 
PG QSOs on the fundamental plane of early-type galaxies lies between the 
loci of moderate-mass and giant Es. It coincides with the region where the 
most massive ULIRGs are typically located, indicating that a subset of PG 
QSOs and ULIRGs have similar dynamical properties.
\item
In some sources, the starburst and strong AGN-accretion 
phase is unambiguously triggered by a merger event. A scenario where 
some ULIRGs may undergo a QSO phase as the merger evolves is therefore
plausible. The same applies for a scenario which assumes that some ULIRGs 
and QSOs may have similar, but not identical, progenitors. Since 
there is no evidence that uniquely favors one of these scenarios, both 
may hold across the PG QSO population. Still, PG QSOs constitute 
only a fraction of the local QSO population, and sources of \mbh\ 
$5 \times 10^8$-10$^9$\msun\ seem to have a different origin. 
\end{enumerate}


\acknowledgments

We are grateful to ESO for the acquisition of excellent-quality data 
during the service-mode observations. K. Dasyra wishes to thank M. 
Vestergaard for 
helpful suggestions. A. Baker acknowledges support from the National 
Radio Astronomy Observatory, which is operated by Associated Universities, 
Inc., under cooperative agreement with the National Science Foundation.
S. Veilleux was supported in part by NASA through grant GO-0987501A.

\clearpage

\clearpage


\begin{deluxetable}{ccccccccc}
\tablecolumns{9}
\tabletypesize{\tiny}
\tablewidth{0pt}
\tablecaption{\label{tab:list} QSO List}
\tablehead{
\colhead{Galaxy} 
& \colhead{RA \tablenotemark{a}} & \colhead{Dec \tablenotemark{b}} 
& \colhead{$z$} & \colhead{$m_B$} 
& \colhead{slit P.A. \tablenotemark{c}}
& \colhead{$t_{\rm integration}$ \tablenotemark{c}}   
& \colhead{$R_{\rm eff}$ \tablenotemark{d}} 
& \colhead{$<\mu_{\rm eff}>$}\tablenotemark{e} \\

\colhead{} & \colhead{(2000)} & \colhead{(2000)} & \colhead{} & \colhead{mag} 
& \colhead{(\degree)}  & \colhead{(mins)} &  \colhead{(kpc)} 
& \colhead{(mag arcsec$^{-2}$)}
}

\startdata
PG 0007$+$106 &00:10:31.0& +10:58:30 & 0.0893 & 16.1 \tablenotemark{f} 
& 139,49 & 160,160 &3.66 &17.1  \tablenotemark{i}\\
PG 0050$+$124 & 00:53:34.9& +12:41:36 & 0.0611 & 14.4 \tablenotemark{f} 
& 134,44 & 80,80 & 1.97  &15.1 \tablenotemark{i} \\
LBQS 0307$-$0101 &03:10:27.8& -00:49:51 & 0.0804 & 16.3 \tablenotemark{g} 
& 44,134 & 120,120 &\nodata &\nodata \\
PG 1119$+$120 &11:21:47.1& +11:44:18 & 0.0502 & 15.4 \tablenotemark{f}
& 89,179 & 80,80 &1.39  &15.4 \tablenotemark{i} \\
PG 1126$-$041 &11:29:16.6& -04:24:08 & 0.0600 & 15.4 \tablenotemark{f}
& 154,64 & 110,105 &4.21 &17.4 \tablenotemark{i} \\
PG 1211$+$143 &12:14:17.7& +14:03:13 & 0.0809 & 14.4 \tablenotemark{h}
& 0,89   & 130,130 & 2.57 & 16.6 \tablenotemark{j}\\
PG 1229$+$204 &12:32:03.6& +20:09:29 & 0.0603 & 15.5 \tablenotemark{h}
& 29,119 & 140,110 &5.12 &17.6 \tablenotemark{i} \\
PG 1404$+$226 &14:06:21.8& +22:23:46 & 0.0980 & 16.5 \tablenotemark{h} 
& 45,135 & 120,120 & 7.15  & 18.5 \tablenotemark{j} \\
PG 1426$+$015 &14:29:06.6& +01:17:06 & 0.0865 & 15.7 \tablenotemark{h} 
& 60,-30 & 180,120 & 6.49 & 17.3 \tablenotemark{j} \\
PG 1617$+$175 &16:20:11.3& +17:24:28 & 0.1124 & 15.5 \tablenotemark{h} 
& -1,89  & 120,110 & 2.43 & 16.5 \tablenotemark{j} \\
PG 2130$+$099 &21:32:27.8& +10:08:19 & 0.0630 & 14.7 \tablenotemark{h}
& -41,49 & 150,70 & 4.44  &17.6 \tablenotemark{i} \\
PG 2214$+$139 &22:17:12.2& +14:14:21 & 0.0658 & 15.0 \tablenotemark{f}
& -1,89 & 120,120 & 3.97 & 17.4 \tablenotemark{j} \\

\enddata

\tablenotetext{a,b}
{Units of right ascension are hours, minutes and seconds, and units of 
declination are degrees, arcminutes, and arcseconds.}

\tablenotetext{c}
{Slit position angles and respective integration times per slit.}

\tablenotetext{d}
{The effective radii are from \cite{mcleod94a} and \cite{veilleux06}.}

\tablenotetext{e}
{The $H-$band mean surface brightness within the effective radius. }

\tablenotetext{f}
{Apparent magnitudes from \cite{pg}.}

\tablenotetext{g}
{Value taken from \cite{chaffee}. Sloan Digital Sky Survey data indicate
a $G-$ and an $R-$ band magnitude of 15.92 and 15.77 respectively. An 
interpolation between these two values yields an expected $B$-band magnitude 
somewhat smaller than that of \cite{chaffee}.}

\tablenotetext{h}
{The (epoch-averaged) $B$-band magnitude is from \cite{kaspi00}. }

\tablenotetext{i}
{The $K$-band host surface brightness is computed from NICMOS $H$-band 
magnitudes (\citealt{veilleux06}) using an $H-K$ color index of 0.3 mag 
(\citealt{jarrett}).}

\tablenotetext{j}
{The $K$-band host surface brightness is computed from NICMOS $H$-band 
magnitudes (\citealt{mcleod94a}) using an $H-K$ color index of 0.3 mag 
(\citealt{jarrett}).}



\end{deluxetable}

\clearpage

\begin{deluxetable}{cccccccc}
\tablecolumns{8}
\tabletypesize{\tiny}
\tablewidth{0pt}
\tablecaption{\label{tab:lum} Source IR fluxes and luminosities}
\tablehead{
\colhead{Galaxy} 
& \colhead{log($\lambda L_{\lambda}[5100\dot{A}]/$\hbox{L$_{\odot}$}) 
\tablenotemark{a}} 
& \colhead{ $f(12 \micron)$ \tablenotemark{b}}
& \colhead{ $f(25 \micron)$ \tablenotemark{b}}
& \colhead{ $f(60 \micron)$ \tablenotemark{b}}
& \colhead{ $f(100 \micron)$ \tablenotemark{b}}
& \colhead{log($L_{\rm IR}/$\hbox{L$_{\odot}$}) \tablenotemark{c}} 
& \colhead{$L_{\rm IR}/L_{\rm bbb}$ \tablenotemark{d}}\\

\colhead{} & \colhead{} & \colhead{$mJy$} & \colhead{$mJy$} 
& \colhead{$mJy$} & \colhead{$mJy$} & \colhead{} & \colhead{} 
}
\startdata
PG 0007$+$106 & 11.23 &  83\tablenotemark{e}  & 163  & 192\tablenotemark{e}  
& 221\tablenotemark{f} & 11.34 & 0.38 \\
PG 0050$+$124 & 11.21 & 549  & 1097 & 2293 & 2959 & 11.93 & 1.77 \\
LBQS 0307$-$0101 & \nodata & \nodata & \nodata & \nodata & \nodata & \nodata 
& \nodata \\
PG 1119$+$120 & 10.55 & 120  & 280  & 546  & 746  & 11.13 & 0.72 \\
PG 1126$-$041 & 10.80 & 104  & 309  & 669  & 1172 & 11.34 & 0.69 \\
PG 1211$+$143 & 11.17 & 166\tablenotemark{e}  & 331\tablenotemark{e}  & 
412\tablenotemark{e} & 689  & 11.58 & 0.41 \\
PG 1229$+$204 & 10.50 & 117  & 230\tablenotemark{e}  & 202\tablenotemark{e}  
& 317\tablenotemark{f} & 11.13 & 0.66 \\
PG 1404$+$226 & 10.80 & 26\tablenotemark{f} & 62\tablenotemark{f}  
& $<$154 & 123\tablenotemark{f}  & $<$11.07\tablenotemark{h} &\nodata \\
PG 1426$+$015 & 11.14 & 130\tablenotemark{e}  & 221\tablenotemark{e}  & 
318  & 350\tablenotemark{f} & 11.50 & 0.95 \\
PG 1617$+$175 & 10.90 & 65\tablenotemark{g}  & 71\tablenotemark{g} & 
102\tablenotemark{g} & 544\tablenotemark{g}  & 11.43 & 0.40 \\
PG 2130$+$099 & 10.88 & 186\tablenotemark{e}  & 357\tablenotemark{e}  & 
480\tablenotemark{e}  & 485\tablenotemark{f} & 11.39 & 0.59 \\
PG 2214$+$139 & 11.08 & 61  & 95  & 337  & $<$282 & 
$<$11.04\tablenotemark{h} & 0.32                  \\
\enddata

\tablenotetext{a}
{Optical luminosities taken from \cite{peterson04} or \cite{veste06}.}

\tablenotetext{b}
{All data are from \cite{sanders89} unless otherwise noted.}

\tablenotetext{c}
{Infrared luminosities calculated using the \cite{sami96} prescription
from the MIR fluxes. }

\tablenotetext{d}
{Ratio of integrated IR over big blue bump luminosity (see text for definition)
are derived from \cite{surace01} and \cite{guyon06}.}

\tablenotetext{e}
{Mean of the \cite{sanders89} and \cite{haas03} flux measurement.}

\tablenotetext{f}
{Infrared Space Observatory flux from \cite{haas03}.}

\tablenotetext{g}
{Infrared Astronomical Satellite flux from \cite{IRASZ}.}

\tablenotetext{h}
{Lower limit for log($L_{IR}/$\hbox{L$_{\odot}$}) of PG 1404$+$266 is 10.89, 
when assigning $f(60 \micron)$=0. Lower limit for 
log($L_{IR}/$\hbox{L$_{\odot}$}) of PG 2214$+$139 is 10.99, when assigning 
$f(100 \micron)$=0.}

\end{deluxetable}

\clearpage

\begin{deluxetable}{ccccccccc}
\tablecolumns{9}
\tabletypesize{\tiny}
\tablewidth{0pt}
\tablecaption{\label{tab:masses} Host dispersions, black hole masses and
Eddington efficiencies}
\tablehead{
\colhead{Source} & \colhead{$\sigma$}   & \colhead{apert.\tablenotemark{a}} &
\colhead{template} &
\colhead{$M_{\rm BH}$  \tablenotemark{b}} & 
\colhead{$M_{\rm BH}$(Edd.)   \tablenotemark{c}} & 
\colhead{$\eta_{\rm Edd}$ \tablenotemark{d}} & 
\colhead{$M_{\rm BH}$(rev.)  \tablenotemark{e}} &
\colhead{$M_{\rm BH}$(S.E.)  \tablenotemark{f}}\\

\colhead{(PG)}& \colhead{(km s$^{-1}$)}& \colhead{(\arcsec)} & 
\colhead{star} & \colhead{(\msun)} & \colhead{(\msun)} & \colhead{} & 
\colhead{(\msun)} &  \colhead{(\msun)}
}
\startdata
PG 0007$+$106 & 201 ($\pm$ 61) & 0.22-1.10 & M0III & $1.38\times 10^8$ 
&  $4.05\times 10^7$ & 0.29 & \nodata & $5.35\times 10^8$\\
PG 0050$+$124 & 188 ($\pm$ 36) & 0.22-0.96 & K0I & $1.05 \times 10^8$ 
&  $3.85\times 10^7$ & 0.37 & \nodata & $2.76\times 10^7$ \\
LBQS 0307$-$0101 & 207 ($\pm$ 49) & 0.22-0.96 & M0III & $1.55\times 10^8$ & 
\nodata & \nodata & \nodata & \nodata \\
PG 1119$+$120 & 162 ($\pm$ 28) & 0.22-0.96 & M0III & $5.79\times 10^7$ 
&  $8.39\times 10^6$ & 0.14 & \nodata & $2.95\times 10^7$ \\
PG 1126$-$041 & 194 ($\pm$ 29) &  0.22-1.10 & K0I+M0III & $1.19\times 10^8$ 
&  $1.50\times 10^7$ & 0.13 & \nodata & $5.61\times 10^7$ \\
PG 1211$+$143 & \nodata & \nodata & \nodata & \nodata & $3.48\times 10^7$ &
\nodata & $9.14\times 10^7$ & \nodata \\
PG 1229$+$204 & 162 ($\pm$ 32) & 0.22-1.25 & K0I+M0III & $5.79\times 10^7$ &  
$7.44\times 10^6$ & 0.13 & $7.32\times 10^7$ & 
$1.38\times 10^8$ \\
PG 1404$+$226 & 237 ($\pm$ 52) & 0.22-0.96 & K0I & $2.67\times 10^8$ &  
$1.48\times 10^7$ & 0.06 & \nodata & $7.74\times 10^6$ \\
PG 1426$+$015 \tablenotemark{g} & 185 ($\pm$ 67) & 0.22-1.10 & M0III & 
$9.87\times 10^7$ &  $3.25\times 10^7$ & 0.33 & $1.30\times 10^9$ & 
$1.16\times 10^9$ \\
PG 1617$+$175 & 183 ($\pm$ 47) & 0.37-1.40 & M0III & $9.45\times 10^7$ &  
$1.87\times 10^7$ & 0.20 & $5.94\times 10^8$ & 
$6.76\times 10^8$ \\
PG 2130$+$099 & 172 ($\pm$ 46) & 0.22-0.96 & K0I+M0III & $7.36\times 10^7$ &  
$1.79\times 10^7$ & 0.24  & $4.57\times 10^8$ & 
$8.05\times 10^7$ \\
PG 2214$+$139 & 156 ($\pm$ 18) & 0.37- 1.40 & K0I+M0III & $4.97\times 10^7$ &  
$2.84\times 10^7$ & 0.57 & \nodata & $3.56\times 10^8$ \\
\enddata

\tablenotetext{a}
{The single-sided aperture within which $\sigma$ was measured.}

\tablenotetext{b}
{Dynamical black hole masses estimated from their relation to the bulge 
dispersion (\citealt{tremaine02}).}

\tablenotetext{c}
{Eddington black hole mass, calculated by attributing $\it{all}$ of 
L$_{\rm bol}$ to the AGN (L$_{Edd}$).}

\tablenotetext{d}
{Ratio of Eddington over dynamical black hole mass.}

\tablenotetext{e}
{Reverberation black hole masses from \cite{peterson04}.}

\tablenotetext{f}
{Single-epoch, virial black hole masses from \cite{veste06} with optical
luminosities from \cite{boroson}.}


\tablenotetext{g}
{The velocity dispersion of the SW (stellar-light-dominated) component of
this interacting system is 154 ($\pm$ 27) \kms ; it is derived using 
the M0III giant.}
\end{deluxetable}

\clearpage


\begin{figure*}
\centering
\includegraphics[height=12cm,width=16cm]{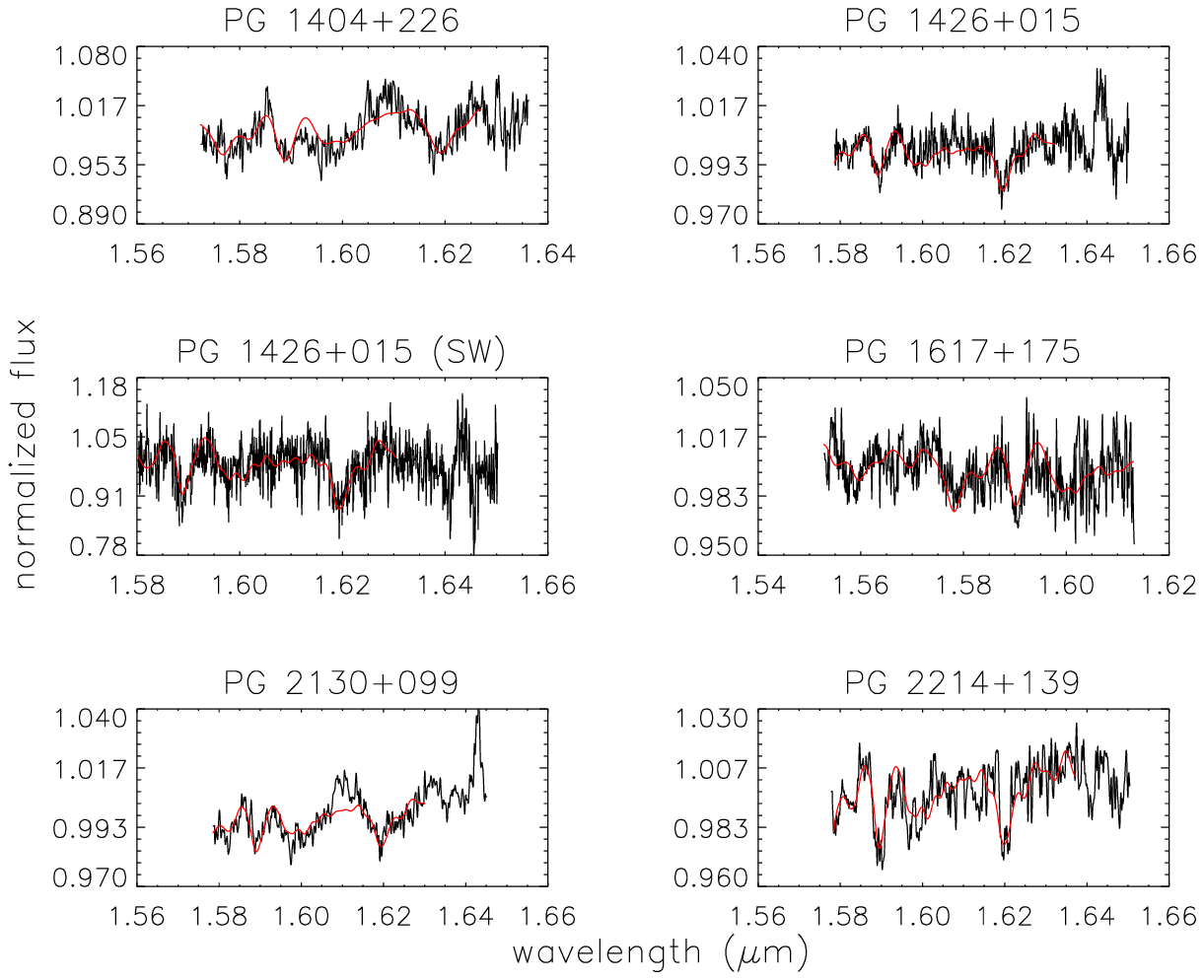}
Fig.~\ref{fig:spectra}. --- continued.
\end{figure*}
\newpage


\end{document}